\documentclass[aps,prb,floatfix,twocolumn,superscriptaddress]{revtex4-2}

\usepackage{mathrsfs}
\usepackage{graphicx}
\usepackage{amsmath}
\usepackage{amssymb}
\usepackage{amsthm}
\usepackage{tikz-cd}
\usepackage{bm}
\usepackage{hyperref}
\usepackage{capt-of}
\usepackage[english]{babel}
\usepackage{cancel}
\usepackage[normalem]{ulem}

\usepackage{epstopdf}
\usepackage{relsize}
\usepackage{CJK}
\usepackage{textcomp}
\usepackage{dsfont}
\usepackage{float}
\renewcommand{\parallel}{\mathrel{/\mskip-4mu/}}

\newcommand{\me}{\mathrm{e}}
\newcommand{\mi}{\mathrm{i}}
\newcommand{\dif}{\mathrm{d}}

\begin{document}
\title{Quantum Geometric Tensor in Parameter-Dependent Quasi-Hermitian Systems}

\author{Ming-Zhang Wang}
\affiliation{School of Physics, Southeast University, Nanjing 211189, China}

\author{Xu-Yang Hou}
\email{houxuyangwow@seu.edu.cn}
\affiliation{School of Physics, Southeast University, Nanjing 211189, China}

\author{Hao Guo}
\email{guohao.ph@seu.edu.cn}
\affiliation{School of Physics, Southeast University, Nanjing 211189, China}
\affiliation{Hefei National Laboratory, University of Science and Technology of China, Hefei 230088, China}

\begin{abstract}
A parameter-dependent physical inner product makes the comparison of quantum states at different parameter points non-unique. We formulate two geometric prescriptions for this comparison in quasi-Hermitian systems. The Type I QGT is defined by an inner-product-compatible covariant derivative and horizontal projection. It is a Hermitian, positive-semidefinite Gram tensor, and, for any normalized solution of the inner-product-preserving dynamics, its tangential component equals the physical energy variance. The Type II QGT is constructed from the endpoint overlap of physical dual states and is invariant under local $GL(1,\mathbb{C})$ rescalings. We show that, on a normalized section, it coincides with the established left--right (LR) QGT, while providing an endpoint-overlap interpretation of that tensor. Our central result is an exact relation between the two constructions: in the positive Dyson frame, the Type I QGT is the positive Gram block of the dressed-derivative sector, while the real symmetric part of the Type II QGT is this metric minus a positive-semidefinite contribution from the metric connection; the full complex tensor also contains cross terms. This relation clarifies the origin of the possible indefiniteness of the Type II metric and separates it from the positivity of the physical Hilbert-space metric. We further distinguish instantaneous Dyson mappings from the fixed inner-product-preserving dynamics and characterize the conditions under which the corresponding QGTs agree. Three two-level models illustrate these results. The physical content of the two tensors is then identified through the energy uncertainty and evolution speed for Type I, and the symmetrized endpoint overlap, complex holonomy, and signed dynamical response for Type II.
\end{abstract}

\maketitle

\section{Introduction}
\label{intro}

The quantum geometric tensor (QGT) characterizes the local geometry of quantum states in parameter space~\cite{QGTCMP80,BRODY200119,QGT10,KOLODRUBETZ20171}. For pure states in the conventional Hermitian setting, its real part is the Fubini--Study metric~\cite{EGUCHI1980213} measuring the distance between physical states, and its imaginary part is the Berry curvature~\cite{Berry84,Simon83}, linking local geometry to global topology~\cite{PhysRevLett.99.100603,PhysRevLett.117.045303,PhysRevB.90.165139,PhysRevA.92.063627,PhysRevB.105.045144,PhysRevLett.121.170401,PhysRevB.104.045103}. This unified structure makes the QGT essential in quantum information, condensed matter, and AMO physics~\cite{IG_Book,Bohm03,KOLODRUBETZ20171,QGTCMP80,cmp/1103904831,RevModPhys.82.1959,PhysRevResearch.3.L042018,PhysRevB.74.085308,PhysRevLett.72.3439,PhysRevB.103.014516,PhysRevB.103.205415,PhysRevB.102.155407,Park26,Resta26} (see Refs.~\cite{Yu25QGT,Gao26} for a review). It has been measured on platforms including nitrogen-vacancy centers~\cite{10.1093/nsr/nwz193}, superconducting qubits~\cite{PhysRevLett.122.210401}, photoemission spectroscopy~\cite{Kang2025}, and ultracold atoms~\cite{Yi23}, and applied to response functions and topological indicators~\cite{PhysRevB.108.094508,PhysRevB.87.245103,PhysRevB.97.201117,PhysRevB.97.041108,PhysRevLett.121.020401,PhysRevB.97.195422,PhysRevLett.124.197002,PhysRevX.10.041041}, quantum phase transitions~\cite{PhysRevLett.99.100603,Zanardi2007}, orbital magnetic susceptibility~\cite{PhysRevB.91.214405,PhysRevB.94.134423}, and superfluidity in lattice systems~\cite{PhysRevLett.117.045303}.

Non-Hermitian Hamiltonians have uncovered phenomena including Anderson localization~\cite{PhysRevLett.77.570}, unconventional quantum emitters~\cite{GZ22,Roccati_2022}, and distinctive topology~\cite{PhysRevX.8.031079,PhysRevLett.121.026808,Roccati2023}, with effects such as the skin effect and novel bulk-boundary correspondences~\cite{PhysRevLett.121.086803,Li2019NatCommun,PhysRevLett.123.230401,doi:10.1126/science.aaw8205,PhysRevLett.126.083604,Xiao2020NatPhys,PhysRevLett.125.126402}, and recently a hidden (2+1)-dimensional electrodynamics in pseudo-Hermitian two-level systems~\cite{kcls-cn82}. The geometric phase has been extended to non-Hermitian systems~\cite{JPAGW13} and used to construct the QGT~\cite{PhysRevA.99.042104}. Among them, quasi-Hermitian systems possess real spectra and admit a positive-definite metric operator $\eta$~\cite{RevModPhys.15.175,PhysRevLett.80.5243,Mostafazadeh2002I,Mostafazadeh2002II,Mostafazadeh2002III}, with $\mathcal{PT}$-symmetric systems as a major branch~\cite{PhysRevLett.80.5243,PhysRevB.82.052404,PhysRevLett.110.083604,Korff_2007,Korff_2008,RevModPhys.88.035002,Bender1999} realized in acoustics, optics, electronics, and quantum systems~\cite{Cham_2015,Feng_2017}. Their dynamics and geometry are defined by the physical inner product $(\cdot,\cdot)_\eta=\langle\cdot|\eta|\cdot\rangle$, generally parameter-dependent~\cite{doi:10.1142/S0219887810004816,Das_2011}, which fundamentally alters the Hilbert-space geometry.

Extending the QGT to quasi-Hermitian systems requires care, since metric variation enters the norm derivative and the evolution equation must be modified. A dynamical framework for time-dependent $\mathcal{PT}$-symmetric systems was established, including proper transformations mapping the corrected evolution to the conventional Schr\"odinger equation~\cite{JPAGW13}, and the link between dynamics and geometry for time-dependent Hilbert spaces was analyzed, distinguishing the evolution generator, energy observable, and representations~\cite{Mostafazadeh2020TD}. An extended QGT was constructed for $\mathcal{PT}$-symmetric systems, unifying the quantum metric and Berry curvature and characterizing quantum phase transitions and spontaneous $\mathcal{PT}$ breaking~\cite{Zhang2019PRA}; a later study extended this to time-dependent states and showed that the induced metric can be pseudo-Riemannian~\cite{Zhang2021CPB}. The positive-definite inner-product operator and the metric induced on parameter space by a specific QGT are distinct, and different QGT constructions need not share the same positivity.

Recent work has examined Hilbert-space bundle geometry and parallel transport~\cite{Ju2024Parallel}, and covariant Berry connections for parameter-dependent metrics~\cite{Arkhipov2026}. In particular, Das \emph{et al.} introduced a left--right (LR) QGT and decomposed it into dressed-derivative and metric-stretching contributions~\cite{Das2026}. On a normalized section, this LR tensor coincides algebraically with the Type II tensor studied here. We therefore do not claim a new algebraic LR-QGT; instead, we identify its symmetric endpoint-overlap construction and establish its exact relation to the Type I QGT. The existence and integrability of proper Dyson mappings over multidimensional parameter spaces, including the associated curvature and holonomy obstructions, have been analyzed in our recent work~\cite{Wang2026PRBproper}. Here we recall only the pathwise and common-frame conditions needed to compare the QGTs. The remaining distinction is that the physical dual $\langle\Phi|=\langle\Psi|\eta$ carries the parameter dependence of the metric, so its full derivative defines an endpoint comparison different from the $\eta$-compatible covariant construction.

We formulate two comparison prescriptions for parameter-dependent quasi-Hermitian systems. Type I is defined through the $\eta$-compatible covariant derivative and horizontal projection; we establish its Hermiticity and positivity and determine when it agrees with the Dyson-induced $K$-type QGT. Type II follows from the full derivative of the physical dual and is the endpoint-overlap form of the LR QGT~\cite{Das2026}; we derive its $GL(1,\mathbb{C})$ gauge structure and complex curvature. Our central result is an exact tensor identity: the real symmetric part of Type II equals the Type I metric minus a positive-semidefinite metric-connection term, while the full complex tensor also contains cross terms. This identity identifies the origin of possible negative directions in the Type II metric without compromising positivity of the physical Hilbert-space metric. Three two-level models illustrate the resulting local geometries and their relation to proper Dyson mappings.

The rest of the paper is organized as follows. Section~\ref{Sec2} reviews the Hermitian QGT and its projective and bundle geometry. Section~\ref{Sec3} introduces quasi-Hermitian dynamics, constructs the Type I QGT, discusses its relation to the Dyson-induced $K$-type QGT and proper mappings, and develops the Type II endpoint construction. The relation between Type II and the LR QGT of Ref.~\cite{Das2026}, as well as the exact comparison between the two tensors, is established in the same section. Section~\ref{physical} discusses their physical content: energy uncertainty and evolution speed for Type I, and endpoint overlap, complex holonomy, and signed dynamical response for Type II. Section~\ref{models} presents three two-level models, and Sec.~\ref{Sec6} concludes. The K\"ahler-geometric background is given in Appendix~\ref{app:kahler}.

\section{Overview of Hermitian QGT}\label{Sec2}
Before discussing the QGT in quasi-Hermitian systems, we first review the QGT and its associated geometric structure in Hermitian quantum systems.

We consider a quantum system whose state $|\psi(\boldsymbol{\lambda})\rangle$ is normalized, $\langle\psi|\psi\rangle=1$, in an $N$-dimensional Hilbert space $\mathcal{H}=\mathbb{C}^N$. The state depends smoothly on $k$ real parameters $\boldsymbol{\lambda}=(\lambda^1,\dots,\lambda^k)$, which parametrize the parameter manifold $\mathcal{M}$. The normalized states form the unit sphere $S^{2N-1}=\{|\psi\rangle\in\mathbb{C}^N:\langle\psi|\psi\rangle=1\}$. Physical rays are obtained by identifying states that differ only by a phase, yielding the projective space $\mathbb{CP}^{N-1}=S^{2N-1}/U(1)$.

The raw quantum distance between two nearby states is
\begin{equation}
\dif s^2
=
\big\||\psi(\boldsymbol{\lambda}+\dif\boldsymbol{\lambda})\rangle-|\psi(\boldsymbol{\lambda})\rangle\big\|^2
=
\langle\partial_\mu\psi|\partial_\nu\psi\rangle\dif\lambda^\mu\dif\lambda^\nu,
\end{equation}
where $\partial_\mu=\partial/\partial\lambda^\mu$ and Einstein summation convention is used. This distance is not invariant under the local phase transformation $|\psi\rangle\to \me^{\mi\chi(\boldsymbol{\lambda})}|\psi\rangle$. Decomposing the raw derivative tensor into real and imaginary parts,
$\langle\partial_\mu\psi|\partial_\nu\psi\rangle
=\tilde{g}_{\mu\nu}-\mi\Omega_{\mu\nu}$,
we find that $\tilde{g}_{\mu\nu}$ is symmetric and $\Omega_{\mu\nu}$ is antisymmetric. The antisymmetric part $\Omega_{\mu\nu}$ is gauge invariant, whereas the real part is not. The phase connection is described by the real-valued Berry connection
\begin{equation}
\mathcal{A}_\mu
=
\mi\langle\psi|\partial_\mu\psi\rangle,
\end{equation}
which transforms as $\mathcal{A}_\mu\to\mathcal{A}_\mu-\partial_\mu\chi$ under
$|\psi\rangle\to\me^{\mi\chi}|\psi\rangle$. The gauge-invariant quantum geometric tensor (QGT) is
\begin{equation}
Q_{\mu\nu}
=
g_{\mu\nu}-\mi\Omega_{\mu\nu}
=
\langle\partial_\mu\psi|\partial_\nu\psi\rangle
-
\langle\partial_\mu\psi|\psi\rangle
\langle\psi|\partial_\nu\psi\rangle,
\end{equation}
where $g_{\mu\nu}=\operatorname{Re}Q_{\mu\nu}$ and
$\Omega_{\mu\nu}=-\operatorname{Im}Q_{\mu\nu}$. Thus, the Berry curvature is
$\mathcal{F}_{\mu\nu}=\partial_\mu\mathcal{A}_\nu-\partial_\nu\mathcal{A}_\mu
=-2\operatorname{Im}Q_{\mu\nu}=2\Omega_{\mu\nu}$.
Introducing the projector $P=|\psi\rangle\langle\psi|$, this can be written compactly as
\begin{equation}\label{Qmn}
Q_{\mu\nu}
=
\langle\partial_\mu\psi|(1-P)|\partial_\nu\psi\rangle=\langle\partial_\mu\psi|P^\perp|\partial_\nu\psi\rangle,
\end{equation}
where $P^\perp=1-P$.

Geometrically, the normalized states form the total space $S^{2N-1}$, which is a $U(1)$ principal bundle over the projective space:
\begin{equation}\label{f}
U(1)\hookrightarrow S^{2N-1}\to\mathbb{CP}^{N-1}.
\end{equation}
The base manifold is $\mathbb{CP}^{N-1}$ (physical rays), and the fiber is the $U(1)$ phase. 

The total distance on $S^{2N-1}$ admits a decomposition into a base contribution and a fiber contribution. To see this, write a normalized lift as $|\psi\rangle=\me^{\mi\phi}|\psi_0\rangle$, where $|\psi_0\rangle$ is a chosen local section; the connection one-form $\mathcal{A}=\mathcal{A}_\mu\dif\lambda^\mu$ is evaluated on this section, $\mathcal{A}_\mu=\mi\langle\psi_0|\partial_\mu\psi_0\rangle$. Then
\begin{equation}
\langle\psi|\partial_\mu\psi\rangle\dif\lambda^\mu
=
\langle\psi|\dif\psi\rangle
=
\mi(\dif\phi-\mathcal{A}),
\label{FSm1c3}
\end{equation}
where $\mathcal{A}=\mathcal{A}_\mu\dif\lambda^\mu$. The raw distance on $S^{2N-1}$ is
\begin{equation}\label{dt}
\dif s^2(S^{2N-1})
=
\langle\partial_\mu\psi|\partial_\nu\psi\rangle\dif\lambda^\mu\dif\lambda^\nu.
\end{equation}
The distance on the base $\mathbb{CP}^{N-1}$ is obtained upon projecting out the fiber direction. Using Eq.~(\ref{Qmn}), we have
\begin{align}
 &\dif s^2(\mathbb{CP}^{N-1})=Q_{\mu\nu}\dif\lambda^\mu\dif\lambda^\nu\notag\\
=&
\langle\partial_\mu\psi|\partial_\nu\psi\rangle\dif\lambda^\mu\dif\lambda^\nu
+
(\langle\psi|\partial_\mu\psi\rangle\dif\lambda^\mu)^2.
\label{FSm1c}
\end{align}
Substituting \eqref{FSm1c3} into \eqref{FSm1c} yields
\begin{equation}
\dif s^2(S^{2N-1})
=
\dif s^2(\mathbb{CP}^{N-1})
+
(\dif\phi-\mathcal{A})^2.
\label{FSm1c4}
\end{equation}
This shows that the total distance consists of a base contribution, which is the Fubini--Study distance between physical rays, and a fiber contribution, which measures the phase mismatch. For horizontal (Berry-parallel) transport, $\langle\psi|\dif\psi\rangle=0$, or equivalently $\dif\phi=\mathcal{A}$. Along a closed curve, this condition gives the Berry phase through the accumulated phase $\oint_C\mathcal{A}$; the fiber contribution then vanishes.

\section{QGT of Quasi-Hermitian quantum systems}\label{Sec3}
\subsection{Quasi-Hermitian systems and dynamics}\label{Sec3a}

We first summarize the basic properties and dynamics of a parameter-dependent quasi-Hermitian system. Consider a finite-dimensional non-Hermitian Hamiltonian $H$ depending smoothly on the parameters $\boldsymbol{\lambda}$. The system is quasi-Hermitian if there exists a positive-definite metric operator $\eta=\eta^\dagger>0$ such that $H^\dagger\eta=\eta H$. The physical inner product is defined as
\begin{equation}\label{ipeta}
\langle\phi|\Psi\rangle_\eta=\langle\phi|\eta|\Psi\rangle.
\end{equation}
Assuming that the system has a non-degenerate spectrum, the eigen-equations are
\begin{align}
H|\Psi_n\rangle=E_n|\Psi_n\rangle,\quad
H^\dagger|\Phi_n\rangle=E_n|\Phi_n\rangle,
\end{align}
with $|\Phi_n\rangle=\eta|\Psi_n\rangle$. The eigenstates satisfy the bi-orthonormal relations $\langle\Psi_m|\eta|\Psi_n\rangle=\delta_{mn}$ and the completeness relation $\sum_n|\Psi_n\rangle\langle\Phi_n|=1$. The non-degeneracy assumption is used to introduce this spectral basis; the QGT constructions and dynamical identities below do not require the evolving state to remain an instantaneous eigenstate.

The Hamiltonian $H$ can be related to a Hermitian Hamiltonian $H_0$ by a Dyson mapping. Introducing an operator $\Omega$ such that $\eta=\Omega^\dagger\Omega$, we have
\begin{equation}
H_0=\Omega H\Omega^{-1},\quad H_0=H_0^\dagger.
\end{equation}
The eigenstates of $H$ and $H_0$ are related by $|\Psi_n\rangle=\Omega^{-1}|\psi_n^0\rangle$, where $H_0|\psi_n^0\rangle=E_n|\psi_n^0\rangle$.

For a time-dependent state $|\Psi(t)\rangle$, the ordinary Schr\"odinger equation $\mi\partial_t|\Psi\rangle=H|\Psi\rangle$ does not preserve the physical norm. Indeed, using the quasi-Hermiticity condition, one finds
\begin{equation}
\frac{\dif}{\dif t}\langle\Psi|\eta|\Psi\rangle
=
\langle\Psi|\dot{\eta}|\Psi\rangle,
\end{equation}
which generally does not vanish when $\eta$ depends on time. To preserve the physical norm, we adopt the inner-product-preserving dynamics
\begin{equation}\label{Gd}
\mi D_t^\eta|\Psi\rangle=H|\Psi\rangle,\quad
D_t^\eta=\frac{\partial}{\partial t}+\Gamma_t,\quad
\Gamma_t=\frac{1}{2}\eta^{-1}\dot{\eta}.
\end{equation}
Equivalently, the equation of motion under the ordinary time derivative reads
\begin{equation}\label{Schrodinger_like}
\mi\frac{\partial}{\partial t}|\Psi\rangle
=
\left(H-\mi\Gamma_t\right)|\Psi\rangle
=
\left(H-\frac{\mi}{2}\eta^{-1}\dot{\eta}\right)|\Psi\rangle.
\end{equation}
The equation is chosen precisely because it cancels the unwanted term $\langle\Psi|\dot{\eta}|\Psi\rangle$ in the norm derivative. With the inner-product compatibility condition $\partial_t\eta=\Gamma_t^\dagger\eta+\eta\Gamma_t$, one obtains
\begin{equation}
\frac{\dif}{\dif t}\langle\Psi|\eta|\Psi\rangle
=
\langle D_t^\eta\Psi|\eta|\Psi\rangle
+
\langle\Psi|\eta|D_t^\eta\Psi\rangle
=
0.
\end{equation}
Therefore, the norm-conserving dynamics is governed by $H-\mi\Gamma_t$ rather than by $H$ itself. Note that $\mi\Gamma_t$ is not Hermitian in general; rather, it is anti-$\eta$-Hermitian, $(\mi\Gamma_t)^\dagger\eta=-\eta(\mi\Gamma_t)$. Norm conservation is ensured by the $\eta$-compatibility of the covariant derivative, not by the Hermiticity of the effective generator in the ordinary sense.

An instantaneous Dyson mapping gives a Hermitian representative, but it does not by itself ensure that the transformed state obeys the Schr\"odinger equation generated by that representative. The additional proper condition and its consequences for the QGT are stated in Sec.~\ref{Sec3b}.

\subsection{Type I QGT}\label{Sec3b}

In the Hermitian case, the QGT is the second-order kernel of the overlap between neighboring states, Eq.~\eqref{Qmn}, measuring the distinguishability of infinitesimally separated rays. This presupposes a comparison rule for pairing states at neighboring parameter points. In a quasi-Hermitian system, the physical inner product depends on the parameters, Eq.~\eqref{ipeta}, and the state space is a ray bundle with fiber carrying a $U(1)$ phase or a $GL(1,\mathbb{C})$ complex rescaling. Comparing neighboring rays now requires a connection or equivalent comparison rule, so a QGT is not a single tensor but a family of second-order distinguishability tensors indexed by that rule. We first adopt the $\eta$-compatible connection $\Gamma_\mu=\frac12\eta^{-1}\partial_\mu\eta$, which absorbs metric variation into the covariant derivative and yields a Hermitian, positive semidefinite Gram matrix.

\subsubsection{Quasi-Hermitian $\Gamma$-Type QGT}

With the physical inner product given in Eq.~(\ref{ipeta}), we introduce the $\Gamma$-type covariant derivative as the covariant component form of the time derivative $D_t^\eta$ in Eq.~(\ref{Gd}). The corresponding spatial components read
\begin{equation}
\Gamma_\mu=\frac{1}{2}\eta^{-1}\partial_\mu\eta,
\qquad
D_\mu^\eta=\partial_\mu+\Gamma_\mu.
\label{Gamma_connection}
\end{equation}
This connection satisfies the metric-compatibility relation
\begin{equation}
\partial_\mu\eta=\Gamma_\mu^\dagger\eta+\eta\Gamma_\mu,
\quad
\partial_\mu(\phi,\Psi)_\eta
=
(D_\mu^\eta\phi,\Psi)_\eta
+
(\phi,D_\mu^\eta\Psi)_\eta.
\label{metric_compatible}
\end{equation}
This relation incorporates the metric variation into the differentiation rule, ensuring that covariant comparison is consistent with the physical inner product. 

We consider a smooth normalized pure-state family $|\Psi(\boldsymbol{\lambda})\rangle$. When the family is used to describe spectral geometry, it may be chosen as a non-degenerate eigen-ray branch; the QGT construction itself does not require an instantaneous eigenstate condition. Representatives of the same ray differ by a local $U(1)$ phase. To define the induced geometry under parameter variation, we project out the component of the covariant derivative along the state itself. This is achieved by the physical projector
\begin{equation}
P_\eta=|\Psi\rangle\langle\Psi|\eta,
\quad
P_\eta^\perp=1-P_\eta,
\quad
|h_\mu\rangle=P_\eta^\perp D_\mu^\eta|\Psi\rangle.
\label{horizontal_component}
\end{equation}
The vector $|h_\mu\rangle$ retains only the horizontal component that changes the ray, while the vertical component along $|\Psi\rangle$ is removed.
We now define the $\Gamma$-type covariant QGT, referred to as the Type I QGT, as
\begin{equation}
Q_{\mu\nu}^{(\eta,\Gamma)}
=
\langle D_\mu^\eta\Psi|\eta P_\eta^\perp|D_\nu^\eta\Psi\rangle
=
\langle h_\mu|\eta|h_\nu\rangle.
\label{QGT_eta_definition}
\end{equation}
This construction is the quasi-Hermitian analogue of the Hermitian QGT in Eq.~(\ref{Qmn}). In both cases, the projector removes the component along the state itself, so that the resulting tensor is invariant under local phase transformations. Likewise, $Q_{\mu\nu}^{(\eta,\Gamma)}$ is invariant under $|\Psi\rangle\to\me^{\mi\chi}|\Psi\rangle$. Since $\eta>0$, $Q^{(\eta,\Gamma)}$ is the Gram matrix of the horizontal vectors under the physical inner product, and is thus Hermitian and positive semidefinite. Here Hermiticity means $Q_{\mu\nu}^{(\eta,\Gamma)}=(Q_{\nu\mu}^{(\eta,\Gamma)})^*$, so that its antisymmetric part is entirely determined by its imaginary part,
\begin{equation}
Q_{\mu\nu}^{(\eta,\Gamma)}-Q_{\nu\mu}^{(\eta,\Gamma)}
=
2\mi\operatorname{Im}Q_{\mu\nu}^{(\eta,\Gamma)}.
\end{equation}
Its real and imaginary parts define the quantum metric and the antisymmetric two-form,
\begin{equation}
g_{\mu\nu}^{(\eta,\Gamma)}=\operatorname{Re}Q_{\mu\nu}^{(\eta,\Gamma)},
\qquad
F_{\mu\nu}^{(\eta,\Gamma)}=-2\operatorname{Im}Q_{\mu\nu}^{(\eta,\Gamma)}.
\label{QGT_eta}
\end{equation}
The metric $g^{(\eta,\Gamma)}$ may be degenerate on the parameter space; for instance, if a parameter direction only changes the representation but not the covariant ray, the corresponding line element vanishes.

\subsubsection{Hermitian K-type QGT}

A Dyson mapping relates a quasi-Hermitian system to an equivalent Hermitian system. Since the latter has its own standard QGT, a natural question arises: does the QGT obtained through the Dyson mapping agree with the Type I QGT defined in the original quasi-Hermitian representation? In this subsection we address this question.

At a fixed parameter point $\boldsymbol{\lambda}$, the Dyson decomposition $\eta=\Omega^\dagger\Omega$ gives the Hermitian Hamiltonian and the transformed state
\begin{equation}
H_H=\Omega H\Omega^{-1}=H_H^\dagger,
\quad
|\psi_H\rangle=\Omega|\Psi\rangle,
\label{instantaneous_Dyson}
\end{equation}
satisfying $\langle\psi_H|\psi_H\rangle=1$.
The Hermitian QGT of the transformed state family is
\begin{equation}
Q_{\mu\nu}^{H}[\psi_H]
=
\langle\partial_\mu\psi_H|P^\perp_H|\partial_\nu\psi_H\rangle,
\quad
P_H^\perp=1-|\psi_H\rangle\langle\psi_H|.
\end{equation}
To relate this tensor to the original representation, we note that the ordinary derivative of $|\psi_H\rangle$ is mapped to an $\Omega$-type covariant derivative of $|\Psi\rangle$,
\begin{equation}
\partial_\mu|\psi_H\rangle=\Omega D_\mu^\Omega|\Psi\rangle,
\quad
D_\mu^\Omega=\partial_\mu+K_\mu^\Omega.
\label{D_Omega2}
\end{equation}
where $K_\mu^\Omega=\Omega^{-1}\partial_\mu\Omega$.
Using the identity
\begin{equation}
1-|\psi_H\rangle\langle\psi_H|=(\Omega^{-1})^\dagger\eta P_\eta^\perp\Omega^{-1},
\end{equation}
the associated standard Hermitian QGT can be written as the $K$-type covariant QGT of the quasi-Hermitian system,
\begin{equation}
Q_{\mu\nu}^{H}[\psi_H]
=
\langle D_\mu^\Omega\Psi|\eta P_\eta^\perp|D_\nu^\Omega\Psi\rangle
\equiv Q_{\mu\nu}^{(\eta,\Omega)}.
\label{K_Omega}
\end{equation}
Thus, for any given smooth Dyson mapping, the $K$-type QGT is precisely the pullback of the Hermitian QGT of the transformed state family.

However, the Dyson mapping is not unique. Any parameter-dependent unitary operator $U(\boldsymbol{\lambda})$ gives another decomposition $\Omega'=U\Omega$, whose connection becomes
\begin{equation}
K_\mu^{\Omega'}=K_\mu^\Omega+
\Omega^{-1}U^{-1}(\partial_\mu U)\Omega.
\label{dyson_connection_transformation}
\end{equation}
The additional term generally changes the horizontal derivative and hence $Q^{(\eta,\Omega)}$. This is not a consequence of the overall phase freedom of the state vector, but rather a change of the representation and connection used to compare the parameter-dependent state family. 



\subsubsection{Equivalence via proper Dyson mapping}

The Dyson-induced $K$-type QGT is the pullback of the Hermitian QGT and generally depends on the chosen Dyson mapping. Its relation to the Type I QGT is controlled by the proper condition. For a general smooth mapping, the transformed state obeys
\begin{equation}
\begin{split}
\mi\frac{\dif}{\dif t}|\psi_H\rangle
&=\left[H_H+\mi(\dot{\Omega}-\Omega\Gamma_t)\Omega^{-1}\right]|\psi_H\rangle\\
&=\left[H_H+\mi\Omega(K_t^\Omega-\Gamma_t)\Omega^{-1}\right]|\psi_H\rangle,
\end{split}
\label{dyson_dynamics}
\end{equation}
where $K_t^\Omega=\Omega^{-1}\dot{\Omega}$. The additional term is Hermitian because both derivatives are compatible with the physical inner product. It vanishes as an operator precisely when
\begin{align}
\dot{\Omega}_p=\Omega_p\Gamma_t
&\Longleftrightarrow K_t^{\Omega_p}=\Gamma_t
\notag\\
&\Longleftrightarrow
\Omega_p^\dagger\dot{\Omega}_p
=\frac12\dot{\eta}
=\dot{\Omega}_p^\dagger\Omega_p.
\label{proper_time_condition}
\end{align}
This is the pathwise proper condition. For smooth positive-definite $\eta(t)$, it is a first-order matrix equation and admits a smooth solution along any prescribed path. This pathwise existence should be distinguished from the existence of a common proper frame over a parameter region. Along a path satisfying Eq.~\eqref{proper_time_condition},
\begin{equation}
\begin{split}
\frac{\dif}{\dif t}(\Omega_p|\Psi\rangle)&=\Omega_pD_t^\eta|\Psi\rangle,\\
Q_{tt}^{(\eta,\Gamma)}&=Q_{tt}^{(\eta,\Omega_p)}=Q_{tt}^{H}[\psi_H].
\end{split}
\label{path_qgt_equal}
\end{equation}
Thus a pathwise proper mapping guarantees equality only of the tangential contraction $Q_{tt}=Q_{\mu\nu}\dot\lambda^\mu\dot\lambda^\nu$; it does not imply componentwise equality of the full tensors.

A component-by-component equality requires one smooth Dyson mapping to satisfy the proper condition in every relevant parameter direction,
\begin{equation}
\Omega_p^\dagger\partial_\mu\Omega_p=\frac12\partial_\mu\eta
\quad\Longleftrightarrow\quad
\partial_\mu\Omega_p=\Omega_p\Gamma_\mu.
\label{QGT_eta_Omega}
\end{equation}
Such a mapping is a common proper frame. It gives $K_\mu^{\Omega_p}=\Gamma_\mu$ and hence
\begin{equation}
Q_{\mu\nu}^{(\eta,\Gamma)}
=Q_{\mu\nu}^{(\eta,\Omega_p)}
=Q_{\mu\nu}^{H}[\psi_H].
\label{proper_qgts_equal}
\end{equation}
The local integrability condition for this common frame is the flatness of the metric-compatible connection,
\begin{equation}
\mathcal{R}_{\mu\nu}^\eta
=\partial_\mu\Gamma_\nu-\partial_\nu\Gamma_\mu
+[\Gamma_\mu,\Gamma_\nu]=0.
\label{ambient_curvature}
\end{equation}
A global common frame may additionally require trivial holonomy on non-contractible loops. These local and global conditions specify when the Dyson-induced and Type I QGTs can coincide; their general obstruction theory is discussed in Ref.~\cite{Wang2026PRBproper}.

\subsubsection{Geometrical structure}

The QGT is directly related to the geometry of the parameter-dependent quasi-Hermitian state family: it measures how the physical rays change under parameter variation, once the metric variation has been absorbed into the covariant connection.

\emph{The bundle structure.} For a chosen smooth ray family, in particular a non-degenerate eigen-ray branch when spectral geometry is considered, the normalized representatives $|\Psi_n(\boldsymbol{\lambda})\rangle$ with $\langle\Psi_n|\eta|\Psi_n\rangle=1$ form a $U(1)$ principal bundle over the physical ray space. Denoting the $\eta$-normalized state space by $\mathcal{S}_\eta$ and the physical ray space by $\mathcal{R}_\eta$, we have
\begin{equation}
U(1)\hookrightarrow\mathcal{S}_\eta\to\mathcal{R}_\eta.
\end{equation}
The parameter manifold $\mathcal{M}$ enters via the chosen state family, which defines a map $\varphi:\mathcal{M}\to\mathcal{R}_\eta$ by $\boldsymbol{\lambda}\mapsto[|\Psi_n(\boldsymbol{\lambda})\rangle]$. The relevant bundle over parameter space is the pullback $\varphi^*(\mathcal{S}_\eta)\to\mathcal{M}$, whose fiber is again the $U(1)$ phase factor.
The associated $U(1)$ gauge connection specifies how the phase fiber is parallel transported along the base manifold. In the present setting, this connection is the Berry connection. When pulled back to parameter space, it reads
\begin{align}
\mathcal{A}_{n,\mu}^\eta
&=
\mi\langle\Psi_n|\eta D_\mu^\eta|\Psi_n\rangle\notag\\
&=
\mi\left(
\langle\Psi_n|\eta|\partial_\mu\Psi_n\rangle
+\frac12\langle\Psi_n|\partial_\mu\eta|\Psi_n\rangle
\right).
\label{A_eta}
\end{align}
Under the local phase transformation $|\Psi_n\rangle\to\me^{\mi\chi}|\Psi_n\rangle$, it transforms as $\mathcal{A}_n^\eta\to\mathcal{A}_n^\eta-\dif\chi$.

\emph{The decomposition of distance.} The local geometry of the bundle is reflected in the distance decomposition. In analogy with the Hermitian case, where the total distance on $S^{2N-1}$ is given by Eq.~(\ref{dt}), the total distance on the $\eta$-normalized state space $\mathcal{S}_\eta$ is defined via the physical inner product. Writing the phase-lifted state as $|\Phi_n\rangle=\me^{\mi\theta_n}|\Psi_n\rangle$ with $\theta_n$ real, we have
\begin{equation}
\dif s_{\mathrm{total}}^2
=
\langle D_\mu^\eta\Phi_n|\eta|D_\nu^\eta\Phi_n\rangle\dif\lambda^\mu\dif\lambda^\nu.
\end{equation}
Using the decomposition $D_\mu^\eta\Phi_n=P_n^\perp D_\mu^\eta\Phi_n+P_n D_\mu^\eta\Phi_n$, we obtain
\begin{equation}
\begin{split}
\dif s_{\mathrm{total}}^2&=\dif s_{\mathrm{base}}^2+\dif s_{\mathrm{fiber}}^2,\\
\dif s_{\mathrm{base}}^2&=g_{\mu\nu}^{(\eta,\Gamma)}\dif\lambda^\mu\dif\lambda^\nu,\\
\dif s_{\mathrm{fiber}}^2&=\big|\langle\Phi_n|\eta D_\mu^\eta\Phi_n\rangle\big|^2\dif\lambda^\mu\dif\lambda^\nu
=\big(\dif\theta_n-\mathcal{A}_n^\eta\big)^2.
\end{split}
\label{parallel_line_element_decomposition}
\end{equation}
Here $g_{\mu\nu}^{(\eta,\Gamma)}
=
\operatorname{Re}Q_{\mu\nu}^{(\eta,\Gamma)}
=
\frac{Q_{\mu\nu}^{(\eta,\Gamma)}+\overline{Q_{\nu\mu}^{(\eta,\Gamma)}}}{2}$
is the quantum metric in the quasi-Hermitian setting, i.e., the quasi-Hermitian analogue of the Fubini--Study metric. It reduces to the Fubini--Study metric when $\eta=1$ and to the $\eta$-weighted quantum metric in general. We have also used
\begin{equation}
\langle\Phi_n|\eta D_\mu^\eta\Phi_n\rangle
=
\mi\big(\partial_\mu\theta_n-\mathcal{A}_{n,\mu}^\eta\big),
\end{equation}
with $\mathcal{A}_n^\eta=\mathcal{A}_{n,\mu}^\eta\dif\lambda^\mu$. The base distance is the quantum distance on the physical ray space, while the fiber distance measures the phase mismatch. Parallel transport is defined by the vanishing of the fiber distance,
\begin{equation}
\langle\Phi_n|\eta D_t^\eta\Phi_n\rangle=0
\quad\Longleftrightarrow\quad
\dif\theta_n=\mathcal{A}_n^\eta.
\label{parallel_horizontal_definition}
\end{equation}
This condition removes only the phase-fiber component. For a closed path, the accumulated geometric phase is
\begin{equation}
\gamma_n^\eta(C)=\oint_C \mathcal{A}_n^\eta.
\end{equation}

\emph{Dyson-induced parallel transport.} A Dyson mapping also induces a parallel transport on the Hermitian side. For $|\psi_n\rangle=\Omega|\Psi_n\rangle$, the Hermitian horizontal condition $\langle\phi_n^H|\dot{\phi}_n^H\rangle=0$ pulls back to
\begin{equation}
\langle\Phi_n^\Omega|\eta D_t^\Omega\Phi_n^\Omega\rangle=0,
\qquad
D_t^\Omega=\frac{\dif}{\dif t}+\Omega^{-1}\dot{\Omega}.
\label{parallel_Dyson_horizontal}
\end{equation}
The corresponding phase and Berry connection are
\begin{equation}
\begin{split}
-\mi\dot{\theta}_n^\Omega&=\langle\Psi_n|\eta|\dot{\Psi}_n\rangle
+\langle\Psi_n|\Omega^\dagger\dot{\Omega}|\Psi_n\rangle,\\
\mathcal{A}_{n,\mu}^H&=\mi\langle\psi_n|\partial_\mu\psi_n\rangle
=\mi\langle\Psi_n|\eta D_\mu^\Omega\Psi_n\rangle.
\end{split}
\label{parallel_Dyson_phase}
\end{equation}
Since $K^\Omega$ is also metric compatible, the same orthogonal decomposition holds with $D^\eta$ and $g^{(\eta,\Gamma)}$ replaced by $D^\Omega$ and $g^{(\eta,\Omega)}$. If $\Omega$ is an arbitrary instantaneous Dyson choice, this parallel transport corresponds to the Hermitian state family it defines and need not coincide with the $\Gamma$-type dynamics. If $\Omega_p$ satisfies the proper condition along the path, the two horizontal conditions and the line elements agree along that path.

\emph{Imaginary part vs.\ Berry curvature.} We now compare the antisymmetric two-form associated with the Type I QGT and the Berry curvature. Equation~\eqref{QGT_eta} defines $F_{\mu\nu}^{(\eta,\Gamma)}=-2\operatorname{Im}Q_{\mu\nu}^{(\eta,\Gamma)}$. In the Hermitian limit, this two-form coincides with the Berry curvature. For a parameter-dependent physical metric, the Berry curvature associated with the metric-compatible connection is instead given by
\begin{align}
\mathcal{F}_{n,\mu\nu}^\eta
&=
\partial_\mu\mathcal{A}_{n,\nu}^\eta
-\partial_\nu\mathcal{A}_{n,\mu}^\eta\notag\\
& =
\mi\langle D_\mu^\eta\Psi_n|\eta D_\nu^\eta\Psi_n\rangle
-\mi\langle D_\nu^\eta\Psi_n|\eta D_\mu^\eta\Psi_n\rangle\notag\\
&\quad+\mi\langle\Psi_n|\eta[D_\mu^\eta,D_\nu^\eta]\Psi_n\rangle\notag\\
&=
-2\operatorname{Im}Q_{n,\mu\nu}^{(\eta,\Gamma)}
+\mi\langle\Psi_n|\eta\mathcal{R}_{\mu\nu}^\eta|\Psi_n\rangle,
\label{berry_ambient_relation}
\end{align}
where $\mathcal{R}_{\mu\nu}^\eta=[D_\mu^\eta,D_\nu^\eta]$ is the curvature of the metric-compatible connection. In deriving this relation, we used $\langle D_\mu^\eta\Psi_n|\eta D_\nu^\eta\Psi_n\rangle=Q_{n,\mu\nu}^{(\eta,\Gamma)}+\mathcal{A}_{n,\mu}^\eta\mathcal{A}_{n,\nu}^\eta$, where the product of Berry-connection components is real. Thus, the two-form $F_{\mu\nu}^{(\eta,\Gamma)}$ associated with the imaginary part of the Type I QGT generally differs from the Berry curvature by the expectation value of the connection curvature. They coincide whenever this curvature contribution vanishes, in particular for a parameter-independent metric.

The topology of the eigen-ray bundle over the parameter manifold $\mathcal{M}$ is determined by the Berry curvature, rather than by $F_{\mu\nu}^{(\eta,\Gamma)}$. For any closed oriented two-cycle $\Sigma\subset\mathcal{M}$, the first Chern number is
\begin{equation}
C_1^{(n)}[\Sigma]=\frac{1}{2\pi}\int_\Sigma\mathcal{F}_n^\eta\in\mathbb{Z}.
\end{equation}
Since $\mathcal{F}_n^\eta$ is the curvature of the $U(1)$ connection on the eigen-ray bundle, this integral is quantized. To express the corresponding relation for the imaginary part of the Type I QGT as an integral of two-forms, define
\begin{equation}
\begin{aligned}
\boldsymbol{\mathcal I}_n^{(\eta,\Gamma)}
&=\frac{1}{2}\operatorname{Im}Q_{n,\mu\nu}^{(\eta,\Gamma)}
\,\dif\lambda^\mu\wedge\dif\lambda^\nu,\\
\boldsymbol{\mathcal C}_n^\eta
&=\frac{\mi}{2}
\langle\Psi_n|\eta\mathcal{R}_{\mu\nu}^\eta|\Psi_n\rangle
\,\dif\lambda^\mu\wedge\dif\lambda^\nu .
\end{aligned}
\end{equation}
Equation~\eqref{berry_ambient_relation} then gives
\begin{equation}
\int_\Sigma\boldsymbol{\mathcal I}_n^{(\eta,\Gamma)}
=
-\frac{1}{2}\int_\Sigma\boldsymbol{\mathcal F}_n^\eta
+\frac{1}{2}\int_\Sigma\boldsymbol{\mathcal C}_n^\eta,
\end{equation}
where $\boldsymbol{\mathcal F}_n^\eta
=\frac{1}{2}\mathcal{F}_{n,\mu\nu}^\eta
\,\dif\lambda^\mu\wedge\dif\lambda^\nu$. The connection-curvature contribution is not generally constrained to be quantized. Thus, the integral of the imaginary part of the Type I QGT alone does not, in general, define a Chern number.

This distinction reflects the relation between the fiberwise geometry of physical rays and the geometry induced on the external parameter manifold. For each fixed parameter value, the positive-definite metric $\eta(\boldsymbol{\lambda})$ identifies the corresponding ray space biholomorphically with $\mathbb{CP}^{N-1}$, which carries its natural K\"ahler structure. The Type I QGT, however, is defined on $\mathcal{M}$ through the covariant horizontal map $X\mapsto |h_X\rangle=P_\eta^\perp D_X^\eta|\Psi\rangle$. A K\"ahler structure induced on $\mathcal{M}$ would require the real part of the Type I QGT to be nondegenerate and positive definite, the image of this map to be invariant under multiplication by $\mi$, and the resulting almost-complex structure to be integrable with a closed associated two-form. In general, these conditions are not automatic, and the pointwise K\"ahler structures need not induce a K\"ahler geometry on the parameter manifold. The correction in Eq.~\eqref{berry_ambient_relation} is governed by $\mathcal{R}_{\mu\nu}^\eta$, the curvature of the metric-compatible comparison connection. A detailed discussion is given in Appendix~\ref{app:kahler}.

\subsection{Type II QGT}\label{Sec3c}

\subsubsection{Endpoint construction}

The Type I QGT was constructed by introducing an inner-product-compatible covariant derivative and projecting out the component along the state. This presupposes a connection that absorbs the metric variation. However, in a quasi-Hermitian system the physical inner product itself depends on the parameters, and one may instead compare two endpoint states directly, each paired with its own physical dual, without invoking any connection in between. This alternative route defines the Type II QGT.

For a smooth nonzero pure-state ray family (with $n$ labeling the chosen family), we introduce
\begin{equation}
\langle\Phi_n|=\langle\Psi_n|\eta,
\quad
N_n=\langle\Phi_n|\Psi_n\rangle>0,
\quad
P_n=\frac{|\Psi_n\rangle\langle\Phi_n|}{N_n}.
\label{GL_dual}
\end{equation}
Here $\langle\Phi_n|$ is the \emph{physical dual} and is distinct from the phase-lifted bra of the previous subsection. We do not require $N_n=1$ at this stage. The same ray admits the complex rescaling $|\Psi_n\rangle\mapsto c|\Psi_n\rangle$ with $c\in\mathbb{C}^*\simeq GL(1,\mathbb{C})$. The physical dual then transforms as $\langle\Phi_n|\mapsto c^*\langle\Phi_n|$, rather than being independently specified as an inverse-rescaled left state.

Since the physical dual contains the metric, its parameter variation must be taken as a full derivative,
\begin{equation}
\partial_\mu\langle\Phi_n|
=
\langle\partial_\mu\Psi_n|\eta
+
\langle\Psi_n|\partial_\mu\eta.
\label{GL_dual_derivative}
\end{equation}
The second term is the essential difference from the fixed-metric case. It cannot be neglected, nor is it equivalent to replacing both the left and right ordinary derivatives by $D^\eta$ simultaneously.

For two neighboring points $\boldsymbol{\lambda}$ and $\boldsymbol{\lambda}'=\boldsymbol{\lambda}+\dif\boldsymbol{\lambda}$, we define the symmetrized endpoint overlap
\begin{equation}
\mathcal{F}_n(\boldsymbol{\lambda},\boldsymbol{\lambda}')
=
\frac{
\langle\Phi_n(\boldsymbol{\lambda})|\Psi_n(\boldsymbol{\lambda}')\rangle
\langle\Phi_n(\boldsymbol{\lambda}')|\Psi_n(\boldsymbol{\lambda})\rangle
}{
N_n(\boldsymbol{\lambda})N_n(\boldsymbol{\lambda}')
}.
\label{GL_overlap}
\end{equation}
The forward and backward overlaps employ the physical duals at the initial and final points, respectively, so that the rescaling factors at the two ends cancel exactly. Because the metrics at the two endpoints differ, $\mathcal{F}_n$ is generally complex and is not assumed to lie between zero and one.

\subsubsection{Gauge-invariant QGT}

Retaining the endpoint construction, we define the Type II QGT directly as
\begin{equation}
Q^{\mathrm{GL}}_{n,\mu\nu}
=
\frac{
\langle\partial_\mu\Phi_n|(1-P_n)|\partial_\nu\Psi_n\rangle
}{
N_n
}.
\label{QGT_2}
\end{equation}
Under a local $GL(1,\mathbb{C})$ rescaling, the derivative produces additional terms along the original ray, which are annihilated by $1-P_n$; hence Eq.~\eqref{QGT_2} is gauge invariant.

Geometrically, $Q^{\mathrm{GL}}$ is the second-order kernel of ray distinguishability under the endpoint comparison: expanding Eq.~\eqref{GL_overlap} gives
\begin{equation}
\mathcal{F}_n
=
1
-
Q^{\mathrm{GL}}_{n,\mu\nu}\dif\lambda^\mu\dif\lambda^\nu
+
O(\|\dif\boldsymbol{\lambda}\|^3).
\label{GL_overlap_expansion}
\end{equation}
Thus $Q^{\mathrm{GL}}$ measures the second-order response of the endpoint overlap as the two rays separate in parameter space, without invoking any connection or parallel transport. This is precisely where it departs from the Type I construction: the metric variation enters through the full derivative of the physical dual, Eq.~\eqref{GL_dual_derivative}, rather than being absorbed into a connection. Consequently, $Q^{\mathrm{GL}}$ is a complex bilinear tensor on the real parameter tangent space but generally not a Gram matrix in the physical inner product; it is neither Hermitian nor positive semidefinite in general.

\emph{Relation to the LR QGT of Das \emph{et al.}} On the normalized section $N_n=1$, the Type II definition is exactly the LR QGT used in Ref.~\cite{Das2026}. To make the identification explicit, choose the positive Dyson map $S=\sqrt{\eta}$ and define the Hermitian-frame state $|\psi_n\rangle=S|\Psi_n\rangle$. The corresponding right and left states are $|R_n\rangle=S^{-1}|\psi_n\rangle=|\Psi_n\rangle$ and $\langle L_n|=\langle\psi_n|S=\langle\Psi_n|\eta=\langle\Phi_n|$. Therefore,
\begin{equation}
Q_{n,\mu\nu}^{\mathrm{GL}}
=
\langle\partial_\mu L_n|
\bigl(1-|R_n\rangle\langle L_n|\bigr)
|\partial_\nu R_n\rangle
\equiv Q_{n,\mu\nu}^{\mathrm{LR}} .
\label{GL_LR_identification}
\end{equation}
Thus, Type II is not a new algebraic version of the LR QGT. Its role here is to provide a symmetric endpoint-overlap construction of that tensor and, crucially, to place it in a direct comparison with the independently defined Type I QGT. The $GL(1,\mathbb{C})$ invariance of the endpoint construction refers to the freedom in choosing a representative of a ray and is distinct from the unitary Dyson-map gauge freedom discussed in Ref.~\cite{Das2026}. This distinction makes it possible to identify which part of the LR geometry is positive and connection-induced, and which part is contributed by metric variation.

The relation to the decomposition in Ref.~\cite{Das2026} can be made more explicit in the Hermitian frame. For $S=\sqrt{\eta}$, write the Dyson connection as $\mathcal{C}_\mu=(\partial_\mu S)S^{-1}$. This Hermitian-frame Dyson connection is distinct from the metric-compatible connection $\Gamma_\mu=\frac12\eta^{-1}\partial_\mu\eta$ used to define Type I in the original representation. Decompose $\mathcal C_\mu$ as
\begin{equation}
\mathcal C_\mu
=\mathsf S_\mu+\mathsf K_\mu,
\quad
\mathsf S_\mu^\dagger=\mathsf S_\mu,
\quad
\mathsf K_\mu^\dagger=-\mathsf K_\mu .
\end{equation}
The Hermitian-frame images of the Type I horizontal state and the metric-connection fluctuation are, respectively, the dressed-derivative and stretching-fluctuation states of Ref.~\cite{Das2026}:
\begin{equation}
\begin{aligned}
S|h_\mu\rangle
&=P_H^\perp(\partial_\mu-\mathsf K_\mu)|\psi_n\rangle
\equiv|\Phi_\mu\rangle,\\
S|k_\mu\rangle
&=P_H^\perp\mathsf S_\mu|\psi_n\rangle
\equiv|\sigma_\mu\rangle .
\end{aligned}
\end{equation}
Accordingly, the exact relation in Eq.~\eqref{two_qgt_exact_relation} becomes
\begin{equation}
Q_{\mu\nu}^{\mathrm{LR}}
=
\langle\Phi_\mu|\Phi_\nu\rangle
-\langle\sigma_\mu|\sigma_\nu\rangle
+\langle\sigma_\mu|\Phi_\nu\rangle
-\langle\Phi_\mu|\sigma_\nu\rangle ,
\end{equation}
which is the dressed-derivative/stretching decomposition of the LR QGT. In this form, the Type I QGT is precisely the positive Gram block $\langle\Phi_\mu|\Phi_\nu\rangle$, now treated as a standalone tensor defined by the metric-compatible connection, whereas the full LR/Type II tensor also contains the stretching and mixed contributions.

To obtain a metric on the parameter manifold, we extract the real symmetric part
\begin{equation}
g^{\mathrm{GL}}_{n,\mu\nu}
=
\operatorname{Re}
\frac{
Q^{\mathrm{GL}}_{n,\mu\nu}+Q^{\mathrm{GL}}_{n,\nu\mu}
}{2},
\quad
\dif s_{n,\mathrm{GL}}^2
=
g^{\mathrm{GL}}_{n,\mu\nu}\dif\lambda^\mu\dif\lambda^\nu.
\label{GL_pseudo_metric_definition}
\end{equation}
Since the Gram positivity of the Type I case is absent, $g^{\mathrm{GL}}$ need not be positive; in any non-degenerate, signature-stable region it defines a pseudo-Riemannian metric. This indefiniteness resides on the parameter space and reflects the endpoint comparison, while the physical metric $\eta$ on the Hilbert space remains positive definite throughout.

Finally, the antisymmetric part of the ordered tensor is a gauge curvature,
\begin{equation}
\mathcal{F}^{\mathrm{GL}}_{n,\mu\nu}
=
\mi\left(
Q^{\mathrm{GL}}_{n,\mu\nu}
-
Q^{\mathrm{GL}}_{n,\nu\mu}
\right),
\end{equation}
rather than the imaginary part of $Q^{\mathrm{GL}}$. In the Type I case, $Q^{(\eta,\Gamma)}$ is Hermitian, so $Q^{(\eta,\Gamma)}_{\nu\mu}=\big(Q^{(\eta,\Gamma)}_{\mu\nu}\big)^*$ and the antisymmetric part reduces to $2\mi\,\mathrm{Im}\,Q^{(\eta,\Gamma)}_{\mu\nu}$, which yields $F=-2\,\mathrm{Im}\,Q^{(\eta,\Gamma)}$. Here $Q^{\mathrm{GL}}$ is generally non-Hermitian, and its imaginary part contains a symmetric component that belongs to the complex symmetric structure of the tensor and does not contribute to $F^{\mathrm{GL}}$. The geometric origin of this curvature is clarified below.

\subsubsection{Complex connection and curvature}

The antisymmetric part of $Q^{\mathrm{GL}}$ defines a two-form $\mathcal{F}^{\mathrm{GL}}$. To identify it as a genuine curvature, we introduce the complex gauge connection
\begin{equation}
a^{\mathrm{GL}}_{n,\mu}
=
\frac{\langle\Phi_n|\partial_\mu\Psi_n\rangle}{N_n},
\qquad
\mathcal{A}^{\mathrm{GL}}_{n,\mu}
=
\mi a^{\mathrm{GL}}_{n,\mu},
\label{GL_connection}
\end{equation}
which is the analogue of the Berry connection for the complex ray bundle. Under the rescaling $|\Psi_n\rangle\mapsto c|\Psi_n\rangle$, it transforms as $a^{\mathrm{GL}}\mapsto a^{\mathrm{GL}}+\dif\ln c$; writing $c=\me^{-\mi X}$ with $X\in\mathbb{C}$ gives $\mathcal{A}^{\mathrm{GL}}\mapsto \mathcal{A}^{\mathrm{GL}}+\dif X$. This transformation law contains both the phase and the modulus freedom, so $\mathcal{A}^{\mathrm{GL}}$ is a $GL(1,\mathbb{C})$ gauge connection rather than a $U(1)$ connection.

Along a parameter path, take the lift $|\Psi_n^{\parallel}\rangle=\me^{\mi\theta_n}|\Psi_n\rangle$ with complex $\theta_n$. The condition for  horizontal lift and the phase equation read
\begin{equation}
\frac{\langle\Phi_n^{\parallel}|\dot{\Psi}_n^{\parallel}\rangle}
{\langle\Phi_n^{\parallel}|\Psi_n^{\parallel}\rangle}=0,
\qquad
\dot{\theta}_n
=
\mathcal{A}^{\mathrm{GL}}_{n,\mu}\dot{\lambda}^\mu.
\label{GL_parallel}
\end{equation}
This condition removes the component of the ordinary derivative along the complex ray. It generally does not preserve the physical norm and should not be confused with the inner-product-compatible parallel transport of the Type I construction.
For a closed path $C$ with a single-valued section, the holonomy is
\begin{equation}
\gamma_n^{\mathrm{GL}}(C)=\oint_C \mathcal{A}_n^{\mathrm{GL}},
\label{GL_holonomy}
\end{equation}
The real part of $\gamma_n^{\mathrm{GL}}$ describes the rotation of the lifted state, while its imaginary part describes the change of its modulus; the latter belongs to the chosen geometric transport and does not by itself imply physical gain or loss.

Taking the exterior derivative of Eq.~\eqref{GL_connection} and using the full derivative of the physical dual, Eq.~\eqref{GL_dual_derivative}, we obtain
\begin{equation}
\mathcal{F}^{\mathrm{GL}}_{n,\mu\nu}
=
\partial_\mu \mathcal{A}^{\mathrm{GL}}_{n,\nu}
-
\partial_\nu \mathcal{A}^{\mathrm{GL}}_{n,\mu}
=
\mi\left(
Q^{\mathrm{GL}}_{n,\mu\nu}
-
Q^{\mathrm{GL}}_{n,\nu\mu}
\right).
\label{GL_curvature}
\end{equation}
The abelian structure of $GL(1,\mathbb{C})$ leaves no commutator term. Since $Q^{\mathrm{GL}}$ is generally non-Hermitian, $F^{\mathrm{GL}}_{n,\mu\nu}$ cannot be reduced to $-2\,\mathrm{Im}\,Q^{\mathrm{GL}}$ as in the Type I case; the symmetric imaginary part of $Q^{\mathrm{GL}}$ belongs to its complex symmetric structure and does not contribute to $F^{\mathrm{GL}}$.

\subsubsection{Geometrical structure}

\emph{The bundle structure.} In contrast to the $U(1)$ bundle over the $\eta$-normalized state space $\mathcal{S}_\eta$ underlying the Type I construction, the Type II QGT corresponds to a $GL(1,\mathbb{C})$ principal bundle
\begin{equation}
GL(1,\mathbb{C})\hookrightarrow\mathcal{S}_{\mathbb{C}}\to\mathcal{R},
\end{equation}
where $\mathcal{S}_{\mathbb{C}}$ consists of the non-zero but otherwise unrestricted state vectors, with no normalization imposed, the fiber is the complex rescaling $|\Psi_n\rangle\mapsto c|\Psi_n\rangle$ with $c\in\mathbb{C}^*$, and $\mathcal{R}$ is the space of complex rays. The fiber is real two-dimensional, spanned by one modulus and one phase direction, and the gauge connection on this bundle is precisely $\mathcal{A}^{\mathrm{GL}}$ defined in Eq.~\eqref{GL_connection}.

\emph{The decomposition of distance.} The total distance on $\mathcal{S}_{\mathbb{C}}$ is defined by the physical inner product,
\begin{equation}
\dif s_{\mathrm{total}}^2
=
\langle\dot{\Psi}_n|\eta|\dot{\Psi}_n\rangle\dif t^2.
\end{equation}
Since $P_n$ in Eq.~\eqref{GL_dual} satisfies $P_n^2=P_n$ and $P_n^\dagger\eta=\eta P_n$, it is an orthogonal projector with respect to the physical inner product, and the total distance decomposes as
\begin{equation}
\dif s_{\mathrm{total}}^2
=
\dif s_{\mathrm{fiber}}^2+\dif s_{\mathrm{base}}^2,
\end{equation}
with
\begin{align}
\dif s_{\mathrm{fiber}}^2
&=
\frac{|\langle\Phi_n|\dot{\Psi}_n\rangle|^2}{N_n}\dif t^2
=
|a^{\mathrm{GL}}_{n,\mu}\dot{\lambda}^\mu|^2\dif t^2,\notag\\
\dif s_{\mathrm{base}}^2
&=
\langle\dot{\Psi}_n|\eta(1-P_n)|\dot{\Psi}_n\rangle\dif t^2.
\end{align}
The fiber distance involves both the phase and the modulus directions, reflecting the enlarged $GL(1,\mathbb{C})$ fiber. Its vanishing defines the parallel transport of Eq.~\eqref{GL_parallel}, which need not preserve the physical norm.

\emph{$Q^{\mathrm{GL}}$ versus the base distance.} Using the full derivative of the physical dual, Eq.~\eqref{GL_dual_derivative}, the Type II QGT splits into two structurally different pieces,
\begin{align}
Q^{\mathrm{GL}}_{n,\mu\nu}
&=
\underbrace{
\frac{1}{N_n}\langle\partial_\mu\Psi_n|\eta(1-P_n)|\partial_\nu\Psi_n\rangle
}_{\text{base distance}}\notag\\
&\quad+
\underbrace{
\frac{1}{N_n}\langle\Psi_n|\partial_\mu\eta(1-P_n)|\partial_\nu\Psi_n\rangle
}_{\text{metric variation}}.
\label{GL_base_plus_metric}
\end{align}
The first term is the quadratic form of the base distance; it is Hermitian and positive semidefinite, so the base geometry of the $GL(1,\mathbb{C})$ bundle is itself positive. The second term originates from the parameter dependence of $\eta$ and is neither Hermitian nor positive. Hence the indefiniteness and non-Hermiticity of $Q^{\mathrm{GL}}$ are entirely carried by the metric-variation term, not by the base geometry.

\emph{The induced metric.} Extracting the real symmetric part according to Eq.~\eqref{GL_pseudo_metric_definition} yields a pseudo-Riemannian metric $g^{\mathrm{GL}}$ in any non-degenerate, signature-stable region. As shown in the next subsection, it is related to the Type I quantum metric by
\begin{equation}
g^{\mathrm{GL}}_{\mu\nu}
=
g^{(\eta,\Gamma)}_{\mu\nu}
-
\operatorname{Re}\langle k_\mu|\eta|k_\nu\rangle,
\quad
|k_\mu\rangle=P_\eta^\perp\Gamma_\mu|\Psi\rangle,
\end{equation}
where the subtracted term is positive semidefinite. Negative directions of $g^{\mathrm{GL}}$ arise when this metric-variation term dominates, so $g^{\mathrm{GL}}$ is a difference of two positive semidefinite quadratic forms and is generally indefinite.

\subsection{Comparison of the two QGTs}\label{Sec3d}

The two QGTs are different geometric prescriptions for the same quasi-Hermitian system. Type I absorbs metric variation into the covariant connection and yields a Hermitian, positive-semidefinite Gram matrix. Type II retains metric variation in the endpoint comparison and is generally non-Hermitian; its real symmetric part need not be positive and defines a pseudo-Riemannian metric only on non-degenerate, signature-stable regions.

\subsubsection{Normalization and the common limit}

A natural starting point is the normalized section $N_n=1$. The residual gauge group then reduces to $U(1)$, but the complex connection does not automatically become a $U(1)$ connection. From the normalization condition,
\begin{align}
2\operatorname{Re} a^{\mathrm{GL}}_{n,\mu}
&=
-\langle\Psi_n|\partial_\mu\eta|\Psi_n\rangle,
\notag\\
\mathcal{A}_{n,\mu}^\eta
&=
\mathcal{A}_{n,\mu}^{\mathrm{GL}}
+
\frac{\mi}{2}\langle\Psi_n|\partial_\mu\eta|\Psi_n\rangle.
\label{GL_reduction_connection}
\end{align}
Reducing the gauge freedom and changing the differential prescription are therefore different operations: normalization alone does not turn $Q^{\mathrm{GL}}$ into $Q^{(\eta,\Gamma)}$. The difference comes from the way the metric variation enters the derivative, not from whether the state is normalized. When $\partial_\mu\eta=0$, the distinction disappears and both reduce to
\begin{equation}
Q_{n,\mu\nu}
=
\langle\partial_\mu\Psi_n|\eta
(1-|\Psi_n\rangle\langle\Psi_n|\eta)
|\partial_\nu\Psi_n\rangle,
\label{fixed_metric_limit}
\end{equation}
which further reduces to the standard pure-state QGT for $\eta=1$. This common limit confirms compatibility with the usual Hermitian theory but does not extend the equivalence to parameter-dependent metrics.

\begin{table*}[ht]
\centering
\begin{tabular}{p{3.3cm}p{5.5cm}p{7.0cm}}
\hline\hline
 & Type I: $Q^{(\eta,\Gamma)}$ & Type II: $Q^{\mathrm{GL}}$ \\
\hline
Object compared & covariant horizontal change of the ray & endpoint overlap of two physical states \\
Metric variation & absorbed into $\Gamma_\mu$ & retained via $\partial_\mu\langle\Phi|$ \\
Gauge group & $U(1)$ & $GL(1,\mathbb{C})$ \\
Hermiticity & Hermitian, positive semidefinite & generally non-Hermitian \\
Real part & positive-semidefinite quantum metric & not necessarily positive; may be indefinite \\
Antisymmetric part & $-2\operatorname{Im}Q^{(\eta,\Gamma)}$ & gauge curvature $F^{\mathrm{GL}}$, not reducible to $\operatorname{Im}Q^{\mathrm{GL}}$ \\
Transport & inner-product compatible, norm preserving & removes the ray component, modulus not preserved \\
Relation to Ref.~\cite{Das2026} & positive Gram block of the dressed-derivative sector & coincides with the LR QGT on a normalized section \\
Physical question & how much the state itself changes & how similar the two endpoint states are \\
\hline\hline
\end{tabular}
\caption{Comparison between Type I and Type II quantum geometric tensors.}
\label{tab:qgt_comparison}
\end{table*}

\subsubsection{Explicit relation on a common section}

To write the difference in a directly verifiable form, we compare the two QGTs on the same normalized section. Let $P=|\Psi\rangle\langle\Psi|\eta$, $P^\perp=1-P$, and define
\begin{equation}
|h_\mu\rangle=P^\perp D_\mu^\eta|\Psi\rangle,
\qquad
|k_\mu\rangle=P^\perp\Gamma_\mu|\Psi\rangle.
\label{two_qgt_horizontal_parts}
\end{equation}
Here $|h_\mu\rangle$ is the covariant horizontal change of the Type I construction, while $|k_\mu\rangle$ is the horizontal component of the metric connection acting on the state. Using $\Gamma_\mu^\dagger\eta=\eta\Gamma_\mu=\partial_\mu\eta/2$, one finds
\begin{equation}
\begin{split}
P^\perp\partial_\mu|\Psi\rangle
&=
|h_\mu\rangle-|k_\mu\rangle,\\
(\partial_\mu\langle\Phi|)P^\perp
&=
(\langle h_\mu|+\langle k_\mu|)\eta.
\end{split}
\label{two_qgt_derivative_split}
\end{equation}
Substituting these into Eq.~\eqref{QGT_2} without any symmetrization gives
\begin{equation}
Q^{\mathrm{GL}}_{\mu\nu}
=
Q^{(\eta,\Gamma)}_{\mu\nu}
-
\langle k_\mu|\eta|k_\nu\rangle
+
\langle k_\mu|\eta|h_\nu\rangle
-
\langle h_\mu|\eta|k_\nu\rangle.
\label{two_qgt_exact_relation}
\end{equation}
The first term is the covariant ray change, the second is the horizontal contribution of the metric variation, and the last two are cross terms forming an anti-Hermitian matrix in $\mu,\nu$. Taking the real symmetric part, the cross terms cancel and
\begin{equation}
g^{\mathrm{GL}}_{\mu\nu}
=
g^{(\eta,\Gamma)}_{\mu\nu}
-
\operatorname{Re}\langle k_\mu|\eta|k_\nu\rangle.
\label{two_metrics_relation}
\end{equation}
For any real tangent vector $v^\mu$, the correction contracts to $\|v^\mu k_\mu\|_\eta^2\ge0$, so $g^{\mathrm{GL}}$ is a difference of two positive semidefinite quadratic forms and can have negative directions even though the physical inner product remains positive definite. Equation~\eqref{two_qgt_exact_relation} also shows that $GL(1,\mathbb{C})$ invariance does not force non-Hermiticity: if the cross terms vanish, $Q^{\mathrm{GL}}$ is Hermitian even when $g^{\mathrm{GL}}$ has negative directions.

\subsubsection{Conditions for coincidence}

If $P^\perp\Gamma_\mu|\Psi\rangle=0$ for all relevant directions, the two QGTs coincide. The fixed-metric case is a special case of this sufficient condition; more generally, a parameter-dependent metric still gives $Q^{\mathrm{GL}}=Q^{(\eta,\Gamma)}$ whenever the metric connection acts only along the chosen ray. This concerns the comparison of two QGTs for a given state family and is distinct from the proper-condition comparison of $D^\eta$ and $D^\Omega$ in Sec.~\ref{Sec3b}. The overall contrast is summarized in Table \ref{tab:qgt_comparison}.

\section{Physical content of the two QGTs}
\label{physical}

The two QGTs encode different ways of comparing quantum states in a parameter-dependent physical Hilbert space. Their physical content is identified through quantities that follow directly from their respective constructions: energy uncertainty and evolution speed for Type I, and a symmetrized endpoint overlap, complex holonomy, and signed response for Type II. In this section, the state $|\Psi(t)\rangle$ is an arbitrary normalized solution of the inner-product-preserving dynamics, not necessarily an instantaneous eigenstate of $H(t)$.

\subsection{Type I QGT: energy uncertainty and evolution speed}

Let $|\Psi(t)\rangle$ be any normalized solution of Eq.~(\ref{Gd}), with $H^\dagger(t)\eta(t)=\eta(t)H(t)$. No instantaneous-eigenstate assumption is made. Along this trajectory, the tangential component of the Type I QGT is exactly the squared energy fluctuation in the physical inner product,
\begin{equation}
Q_{tt}^{(\eta,\Gamma)}
=
(\Delta_\eta E)^2
=
\langle\Psi|\eta H^2|\Psi\rangle
-
\langle\Psi|\eta H|\Psi\rangle^2 .
\label{AA_quasihermitian}
\end{equation}
This is an exact identity. Since $D_t^\eta|\Psi\rangle=-\mi H|\Psi\rangle$, one has $|h_t\rangle=-\mi P_\eta^\perp H|\Psi\rangle$ and
$Q_{tt}^{(\eta,\Gamma)}=\langle h_t|\eta|h_t\rangle=\langle\Psi|\eta H P_\eta^\perp H|\Psi\rangle=(\Delta_\eta E)^2$.
Here $\langle H\rangle_\eta=\langle\Psi|\eta H|\Psi\rangle$, and quasi-Hermiticity makes this expectation value real. Equation~\eqref{AA_quasihermitian} is the quasi-Hermitian Anandan--Aharonov relation. In the units $\hbar=1$ used in Eq.~\eqref{Gd}, the Type I speed of the state ray is $\dif s/\dif t=\Delta_\eta E$; restoring $\hbar$ gives $\dif s/\dif t=\Delta_\eta E/\hbar$. The comparison prescription is fixed here by the metric-compatible connection $\Gamma_\mu$.

\subsection{Type II QGT: endpoint overlap, complex holonomy, and signed response}

\emph{Symmetrized endpoint overlap.} The quantity with direct meaning in the Type II construction is the modulus of the endpoint overlap,
\begin{equation}
S_n(\boldsymbol{\lambda},\boldsymbol{\lambda}')
\equiv
|\mathcal{F}_n|^{1/2}
=
\frac{
|\langle\Phi_n(\boldsymbol{\lambda})|\Psi_n(\boldsymbol{\lambda}')\rangle|
}{
\sqrt{N_n(\boldsymbol{\lambda})N_n(\boldsymbol{\lambda}')}
},
\label{symmetrized_overlap}
\end{equation}
which is the geometric mean of the two directed overlaps, each evaluated with the physical dual at its own endpoint. It is real, symmetric under endpoint exchange, and $GL(1,\mathbb{C})$ invariant. Because the two endpoints carry different metrics, the Cauchy--Schwarz inequality does not apply and $S_n$ is not generally bounded by one. Expanding Eq.~\eqref{GL_overlap_expansion}, we obtain
\begin{equation}
S_n
=
1
-
\frac{1}{2}
g^{\mathrm{GL}}_{n,\mu\nu}
\dif\lambda^\mu\dif\lambda^\nu
+
O(\|\dif\boldsymbol{\lambda}\|^3).
\label{S_expansion}
\end{equation}
Thus, for a real tangent vector $v^\mu$, the local quadratic response is
\begin{equation}
-2\lim_{\epsilon\to0}
\frac{\ln S_n(\boldsymbol{\lambda},\boldsymbol{\lambda}+\epsilon\boldsymbol v)}
{\epsilon^2}
=
g^{\mathrm{GL}}_{n,\mu\nu}v^\mu v^\nu .
\label{signed_endpoint_response}
\end{equation}
We refer to this quantity as a signed endpoint-overlap response rather than a fidelity susceptibility. In contrast to a conventional fidelity susceptibility, it need not be nonnegative: a negative value means that the symmetrized endpoint overlap increases locally. This behavior reflects the fact that the two endpoint states are compared using their own parameter-dependent physical duals, rather than a single common inner product.

\emph{Complex geometric phase.} For a closed path with a single-valued section, the holonomy of Eq.~\eqref{GL_holonomy} is
\begin{equation}
\mathcal{W}_n(C)=\me^{\mi\gamma_n^{\mathrm{GL}}(C)}
=
\me^{\mi\operatorname{Re}\gamma_n^{\mathrm{GL}}}
\me^{-\operatorname{Im}\gamma_n^{\mathrm{GL}}}.
\end{equation}
The real part determines the phase of the complex holonomy, while the imaginary part determines its modulus,
\begin{equation}
|\mathcal{W}_n(C)|
=
\me^{-\operatorname{Im}\gamma_n^{\mathrm{GL}}(C)}.
\end{equation}
This modulus characterizes the chosen $GL(1,\mathbb{C})$ geometric transport and representative; it should not be identified with physical gain or loss without specifying an associated dynamical or measurement protocol.

\emph{Dynamical response along the evolution.} For the same general normalized trajectory, contracting Eq.~\eqref{two_metrics_relation} along the inner-product-preserving evolution and using Eq.~\eqref{AA_quasihermitian} gives
\begin{equation}
\operatorname{Re}Q_{tt}^{\mathrm{GL}}
=
(\Delta_\eta E)^2
-
\|k_t\|_\eta^2,
\qquad
|k_t\rangle=P_\eta^\perp\Gamma_t|\Psi\rangle .
\label{type2_dynamical_rate}
\end{equation}
The signed quadratic response of the endpoint overlap along this trajectory is the energy-fluctuation contribution reduced by the metric-connection contribution. When the latter dominates, $\operatorname{Re}Q_{tt}^{\mathrm{GL}}<0$ and the symmetrized endpoint overlap increases locally, consistently with Eq.~\eqref{S_expansion}. The phase of the symmetrized endpoint overlap $\mathcal{F}_n$ is controlled by the symmetric imaginary part,
\begin{equation}
\arg\mathcal{F}_n
=-\operatorname{Im}Q^{\mathrm{GL}}_{n,(\mu\nu)}
\dif\lambda^\mu\dif\lambda^\nu
+O(\|\dif\boldsymbol{\lambda}\|^3),
\end{equation}
which is a Pancharatnam-type phase associated with the parameter-dependent physical dual.

Since the real symmetric part of Type II need not be positive semidefinite, the Type II tensor does not in general define a quantum distance or an Anandan--Aharonov speed. Its physical role is instead to characterize the endpoint comparison and its complex geometric and dynamical responses.
\section{Examples}
\label{models}

Here we test Eqs.~(\ref{two_qgt_exact_relation}) and (\ref{two_metrics_relation}) with two-level models, comparing the horizontal derivatives, metric signatures, and complex tensor structures of the two QGTs. In what follows, we choose a special Dyson mapping $S=\sqrt{\eta}$ for computational convenience. This defines the $S$ frame for the associated Hermitian system, $|\chi_n\rangle=S|\Psi_n\rangle$. In this frame, the connection emerging from the similarity transformation is
$G_\mu=\frac12[\partial_\mu S,S^{-1}]$, rather than $K_\mu$ or $\Gamma_\mu$, because it absorbs the non-commutativity between $S$ and $\partial_\mu$. We therefore have
\begin{equation}
S D_\mu^\eta|\Psi_n\rangle=(\partial_\mu-G_\mu)|\chi_n\rangle,
\label{Qeta-S-frame}
\end{equation}
where $G_\mu$ is anti-Hermitian. The Type I QGT then takes the form
\begin{equation}
Q_{n,\mu\nu}^{(\eta,\Gamma)}
=
\langle(\partial_\mu-G_\mu)\chi_n|
\left(1-|\chi_n\rangle\langle\chi_n|\right)
|(\partial_\nu-G_\nu)\chi_n\rangle.
\label{Qeta-S-frame-explicit}
\end{equation}

\begin{figure}
	\centering
	\includegraphics[width=0.47\textwidth]{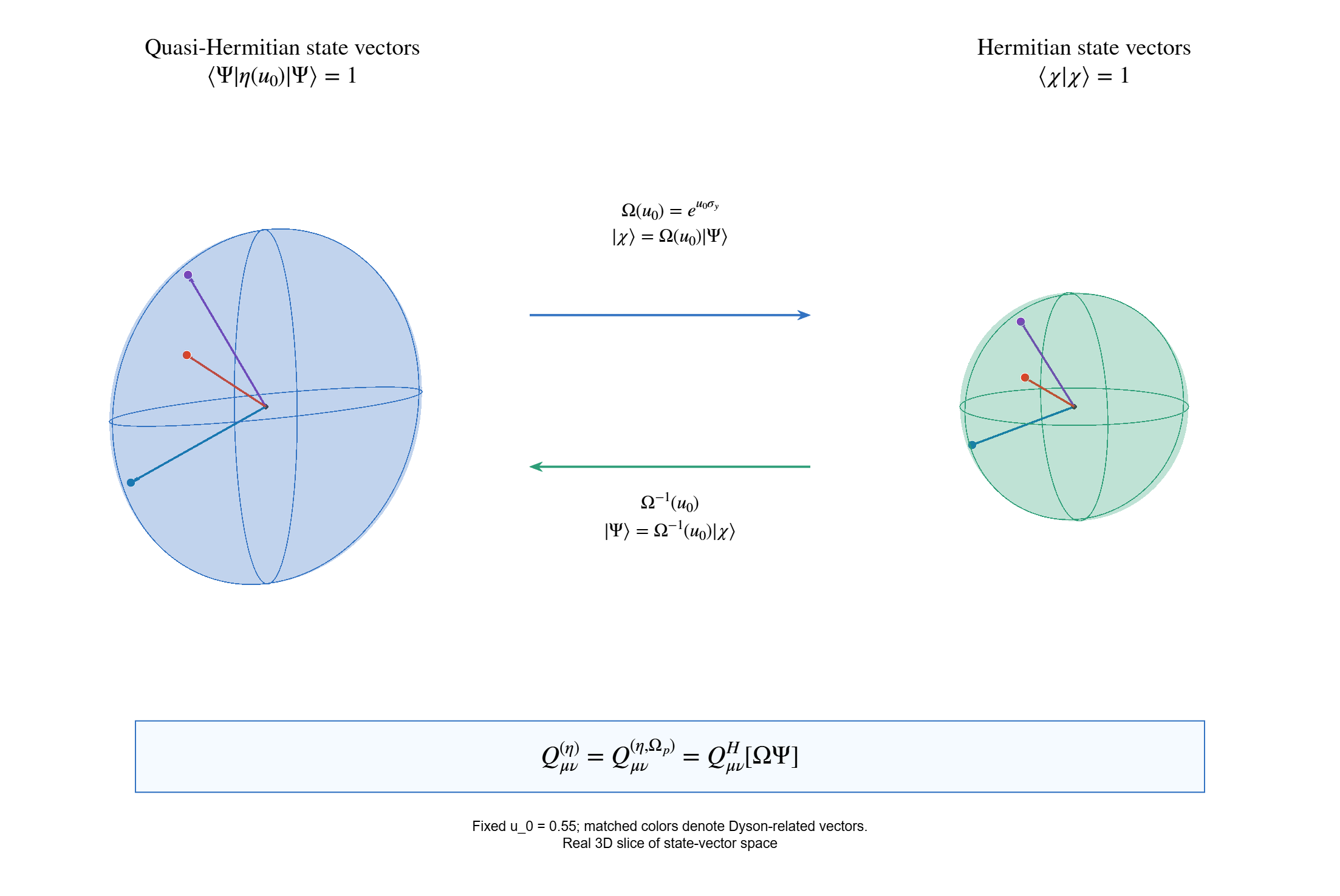}
	\caption{Dyson map for Example 1 with $u_0=0.55$. The ellipsoidal section of the normalized state set and the spherical section are related by $\Omega_p=\me^{u\sigma_y}$. The common proper condition ensures $Q^{(\eta,\Gamma)}=Q^{(\eta,\Omega_p)}=Q^H$.}
	\label{fig:QGT11}
\end{figure}

\subsection{Example 1: separated state and metric parameters}
\label{subex1}

In this model the parameter space is
\begin{equation}
\mathcal{M}=(0,\pi)\times\mathbb{R},
\qquad
\boldsymbol{\lambda}=(\theta,u),
\end{equation}
where $u$ controls the physical metric and $\theta$ controls the eigen-ray in the Hermitian representation, so that the $\eta$-covariant state change and the metric change can be tracked separately. The parameter-dependent metric is
\begin{equation}
\eta(u)=\me^{2u\sigma_y}
=\cosh(2u)\mathbb{I}+\sinh(2u)\sigma_y,
\label{ex1-eta}
\end{equation}
with $S(u)=\sqrt{\eta(u)}=\me^{u\sigma_y}$. Since the eigenvalues $\me^{\pm2u}>0$, the metric is positive definite on the whole parameter space. We choose the Hermitian Hamiltonian $H_H(\theta)=\cos\theta\,\sigma_z+\sin\theta\,\sigma_x$, whose quasi-Hermitian counterpart is
\begin{align}
H(\theta,u)&=S^{-1}(u) H_H(\theta)S(u)\notag\\
&=\left[\sin\theta\cosh(2u)-\mi\cos\theta\sinh(2u)\right]\sigma_x
\notag\\
&+\left[\cos\theta\cosh(2u)+\mi\sin\theta\sinh(2u)\right]\sigma_z.
\label{ex1-H-explicit}
\end{align}
For $u\neq0$, $H(\theta,u)$ is generally non-Hermitian but satisfies $H^\dagger\eta=\eta H$. The positive-energy eigenstate of $H_H$ and its orthogonal partner are
\begin{equation}
|\chi_+(\theta)\rangle=
\begin{pmatrix}
\cos(\frac{\theta}{2})\\
\sin(\frac{\theta}{2})
\end{pmatrix},
\quad
|\chi_-(\theta)\rangle=
\begin{pmatrix}
-\sin(\frac{\theta}{2})\\
\cos(\frac{\theta}{2})
\end{pmatrix}.
\end{equation}
The quasi-Hermitian eigenstate is $|\Psi_+(\theta,u)\rangle=S^{-1}(u)|\chi_+(\theta)\rangle$ with $\langle\Psi_+|\eta|\Psi_+\rangle=1$.

\emph{Covariant connections.} Since $S$ depends only on $u$ and commutes with its derivative, $G_\theta=G_u=0$. In this case $S$ itself satisfies the proper condition, i.e., $S=\Omega_p$, so $Q^{(\eta,\Gamma)}$ can be compared directly with the ordinary Hermitian QGT of $|\chi_+\rangle$. The metric-compatible connection is
\begin{equation}
\Gamma_\theta=0,
\quad
\Gamma_u=\frac12\eta^{-1}\partial_u\eta=\sigma_y,
\label{ex1-Gamma}
\end{equation}
so that
\begin{equation}
D_\theta^\eta|\Psi_+\rangle=S^{-1}\partial_\theta|\chi_+\rangle,
\qquad
D_u^\eta|\Psi_+\rangle=0.
\label{ex1-covariant-derivatives}
\end{equation}
Since $|\chi_+\rangle$ can be chosen real, the quasi-Hermitian Berry connection and curvature vanish,
\begin{equation}
\mathcal{A}_{+,\theta}^{\eta}=\mathcal{A}_{+,u}^{\eta}=0,
\qquad
\mathcal{F}_{+,\theta u}^{\eta}=0.
\label{ex1-Berry}
\end{equation}

\emph{Type I QGT.} From Eq.~(\ref{Qeta-S-frame-explicit}),
\begin{equation}
Q_{+,\mu\nu}^{(\eta,\Gamma)}
=
\begin{pmatrix}
\dfrac14 & 0\\[6pt]
0 & 0
\end{pmatrix}_{(\theta,u)} .
\label{ex1-Qeta}
\end{equation}
Since a common proper frame exists globally,
\begin{equation}
Q_{+,\mu\nu}^{(\eta,\Gamma)}
=
Q_{+,\mu\nu}^{(\eta,\Omega_p)}
=
Q_{+,\mu\nu}^{H}.
\end{equation}
The same $S(u)$ satisfies the proper equation in both parameter directions and is therefore a common proper frame. Figure~\ref{fig:QGT11} illustrates this representation. The vanishing $D_u^\eta|\Psi_+\rangle=0$ shows that the metric variation is fully compensated in the covariant derivative, leaving the degenerate line element $\dif s_\eta^2=\dif\theta^2/4$.

\emph{Type II QGT.} In the $S$ frame, the endpoint construction induces $B_\mu=(\partial_\mu S)S^{-1}$, with $B_\theta=0$ and $B_u=\sigma_y$. Using $|\Psi_+\rangle=S^{-1}|\chi_+\rangle$, $\langle\Phi_+|=\langle\chi_+|S$, and Eq.~(\ref{QGT_2}), one obtains
\begin{equation}
Q_{+,\mu\nu}^{\mathrm{GL}}
=
\begin{pmatrix}
\dfrac14 & -\dfrac{\mi}{2}\\[6pt]
-\dfrac{\mi}{2} & -1
\end{pmatrix}_{(\theta,u)} .
\label{ex1-QGL}
\end{equation}
The Type II QGT is thus nonzero and non-Hermitian. Since it is symmetric in $\mu,\nu$,
\begin{equation}
g_{+,\mu\nu}^{\mathrm{GL}}
=\operatorname{Re}Q_{+,\mu\nu}^{\mathrm{GL}}
=
\begin{pmatrix}
\dfrac14&0\\[6pt]
0&-1
\end{pmatrix}_{(\theta,u)} ,
\end{equation}
with $\dif s_{\mathrm{GL}}^2=\frac14\dif\theta^2-\dif u^2$, a pseudo-Riemannian structure of signature $(1,1)$.

\begin{figure}[ht]
	\centering
	\includegraphics[width=0.47\textwidth]{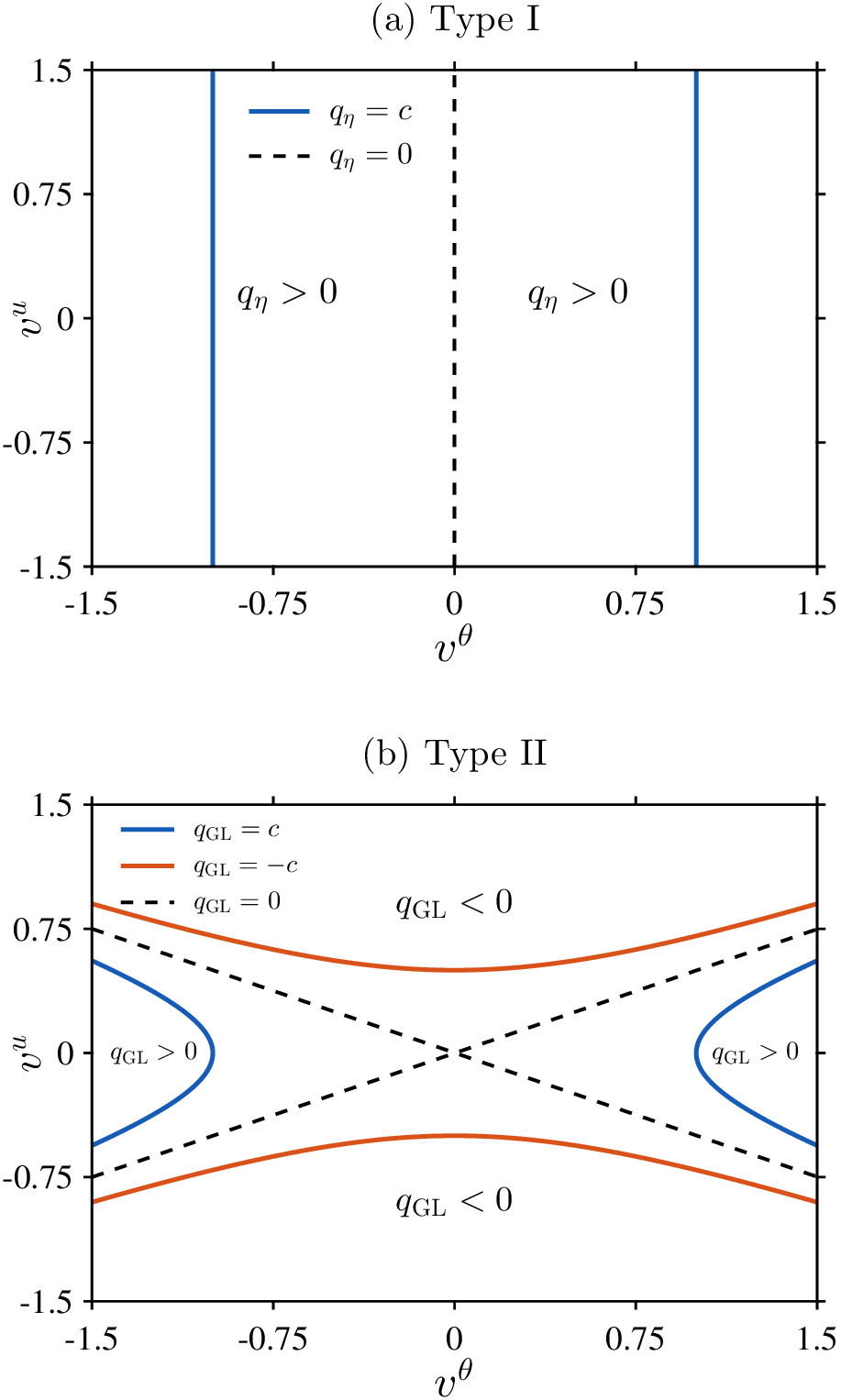}
	\caption{Level sets of the two metrics in the tangent space for Example 1 with $c=1/4$. (a) The Type I metric is positive semidefinite and degenerate along $v^\theta=0$. (b) The Type II metric has signature $(1,1)$ with null directions $v^u=\pm v^\theta/2$.}
	\label{fig:QGT12}
\end{figure}

The decomposition can be checked directly. From $\Gamma_u=\sigma_y$ and $\sigma_y|\chi_+\rangle=\mi|\chi_-\rangle$, one has $|k_\theta\rangle=0$, $|k_u\rangle=\mi S^{-1}|\chi_-\rangle$, and $\langle k_u|\eta|k_u\rangle=1$; Eq.~(\ref{two_metrics_relation}) then gives $g_{uu}^{\mathrm{GL}}=-1$, while the cross terms give $Q_{\theta u}^{\mathrm{GL}}=-\mi/2$. Since $Q_{+,\mu\nu}^{\mathrm{GL}}$ is symmetric in the parameter indices, its antisymmetric part, and hence the associated Berry curvature, vanishes identically. The nonzero imaginary components are therefore purely symmetric and belong to the complex symmetric structure, not to any Berry curvature. Figure~\ref{fig:QGT12} compares $q_\eta(v)=(v^\theta)^2/4$ and $q_{\mathrm{GL}}(v)=(v^\theta)^2/4-(v^u)^2$ in the same tangent space: the zero direction of the former is the kernel of the metric, while the nonzero null vectors of the latter are not in the kernel. Only the real symmetric parts are compared; the nonzero imaginary symmetric components are not represented. This also shows that a common proper frame guarantees the agreement of the two covariant QGTs, not that of $Q^{(\eta,\Gamma)}$ and $Q^{\mathrm{GL}}$.

\emph{Topology.} The parameter space $\mathcal{M}=(0,\pi)\times\mathbb{R}$ is contractible, so the pullback of the $U(1)$ principal bundle over $\mathcal{M}$ is trivial and the first Chern number vanishes,
\begin{equation}
C_1^{(\eta,\Gamma)}=C_1^{\mathrm{GL}}=0.
\end{equation}
This is consistent with the fact that the Berry curvature vanishes identically in this model.

\subsection{Example 2: joint metric variation in two parameters}
\label{subex2}

In this model the radial parameter changes the eigenvalues of the metric, while the angular parameter rotates its eigen-directions. Both variations enter the state vector and its physical dual simultaneously, so that the model exhibits the generic features of $Q^{\mathrm{GL}}$: complex diagonal components, non-Hermiticity, and a real metric that differs from $Q^{(\eta,\Gamma)}$.

We choose the parameter space as the disk $0\le r<1$, $0\le\theta<2\pi$, and the metric $\eta(r,\theta)=\mathbb{I}+rV_\theta$ with $V_\theta=\cos\theta\,\sigma_x+\sin\theta\,\sigma_y$. This metric family also appears in Ref.~\cite{Wang2026PRBproper}. To focus on the QGT comparison, we choose the two-level Hamiltonian $H=H_x\sigma_x+H_y\sigma_y+H_z\sigma_z$ with a radially varying eigenstate; its components are
\begin{align}
H_x&=\sin\beta_r
\left(\frac{\gamma_r+1}{2}\cos\theta
-\frac{\gamma_r-1}{2}\cos3\theta\right)\notag\\
&\quad-\mi\gamma_r r\cos\beta_r\sin\theta,\notag\\
H_y&=\sin\beta_r
\left(-\frac{\gamma_r+1}{2}\sin\theta
-\frac{\gamma_r-1}{2}\sin3\theta\right)\notag\\
&\quad
+\mi\gamma_r r\cos\beta_r\cos\theta,\notag\\
H_z&=\gamma_r\cos\beta_r
+\mi\gamma_r r\sin\beta_r\sin(2\theta),
\label{ex2-H-components}
\end{align}
where $\rho=\sqrt{1-r^2}$, $\gamma_r=\rho^{-1}$, and $\sin\beta_r=r\sin\alpha$, with $\alpha$ a fixed nonzero parameter. This Hamiltonian satisfies $\eta H=H^\dagger\eta$, $H^2=\mathbb{I}$, and $E_\pm=\pm1$. The $E=-1$ eigenstate is
\begin{equation}
|\Psi_-\rangle=\eta^{-1/2}\me^{\mi\theta\sigma_z}|\psi_-\rangle,
\label{ex2-Psi}
\end{equation}
with
\begin{equation}
|\psi_-\rangle
=
\begin{pmatrix}
\sin(\beta_r/2)\me^{-\mi\theta}\\
-\cos(\beta_r/2)
\end{pmatrix}
\label{ex2-psi}
\end{equation}
being its corresponding Hermitian eigenstate. In the $S$ frame, one finds $G_r=0$ and $G_\theta=\mi g(r)\sigma_z$, where $g(r)=\frac{\gamma_r-1}{2}=\frac{1-\rho}{2\rho}$. The state $|\chi_-\rangle$ is given by
\begin{equation}
|\chi_-\rangle=S|\Psi_-\rangle
=\me^{\mi\theta\sigma_z}|\psi_-\rangle
=
\begin{pmatrix}
\sin(\beta_r/2)\\
-\cos(\beta_r/2)\me^{-\mi\theta}
\end{pmatrix},
\end{equation}
with its orthonormal partner
\begin{equation}
|\zeta_-\rangle
=
\begin{pmatrix}
\cos(\beta_r/2)\\
\sin(\beta_r/2)\me^{-\mi\theta}
\end{pmatrix}.
\label{ex2-zeta}
\end{equation}

\emph{Type I QGT.}  According to Eq.~\eqref{Qeta-S-frame},
\begin{equation}
SD_r^\eta|\Psi_-\rangle=\partial_r|\chi_-\rangle,
\;
SD_\theta^\eta|\Psi_-\rangle
=(\partial_\theta-\mi g\sigma_z)|\chi_-\rangle,
\end{equation}
which leads to $\langle\zeta_-|SD_r^\eta|\Psi_-\rangle
=\frac{\beta_r'}2$ and $\langle\zeta_-|SD_\theta^\eta|\Psi_-\rangle
=\mi\kappa(r)\sin\beta_r$ with $\kappa(r)=1-\frac{1}{2\rho}
=\frac{1-2g(r)}{2}$.
The essential feature is that the angular connection $G_\theta=\mi g(r)\sigma_z$ is nonzero, so the horizontal derivative is modified by the connection and the angular component of $Q^{(\eta,\Gamma)}$ acquires the nontrivial weight $\kappa(r)$. This yields
\begin{equation}
Q^{(\eta,\Gamma)}
=
\begin{pmatrix}
\dfrac{(\beta_r')^2}{4}
&
\dfrac{\mi}{2}\beta_r'\kappa(r)\sin\beta_r
\\[6pt]
-\dfrac{\mi}{2}\beta_r'\kappa(r)\sin\beta_r
&
\kappa(r)^2\sin^2\beta_r
\end{pmatrix}_{(r,\theta)} .
\label{ex2-Qeta}
\end{equation}
Its real symmetric part gives the quantum metric
\begin{equation}
\dif s_\eta^2=\frac{(\beta_r')^2}{4}\dif r^2
+\kappa(r)^2\sin^2\beta_r\dif\theta^2.
\label{ex2-metric-line}
\end{equation}
Both coefficients are positive for $0<r<1$, $r\neq\sqrt3/2$; at $r=\sqrt3/2$, $\kappa(r)=0$ degenerates the angular covariant line element. On a fixed circle $r=r_0$, with $\beta_0=\beta_{r_0}$,
\begin{equation}
Q_{\theta\theta}^{(\eta,\Gamma)}
=\kappa(r_0)^2\sin^2\beta_0
=\frac{(2g(r_0)-1)^2}{4}\sin^2\beta_0.
\label{ex2-Qeta-circle}
\end{equation}

For this metric family, the metric-induced connection has nonzero curvature on $0<r<1$, so no common proper frame exists on the two-dimensional parameter region. We do not repeat the pathwise construction here and instead focus on the two QGTs.

\begin{figure}
	\centering
	\includegraphics[width=0.48\textwidth]{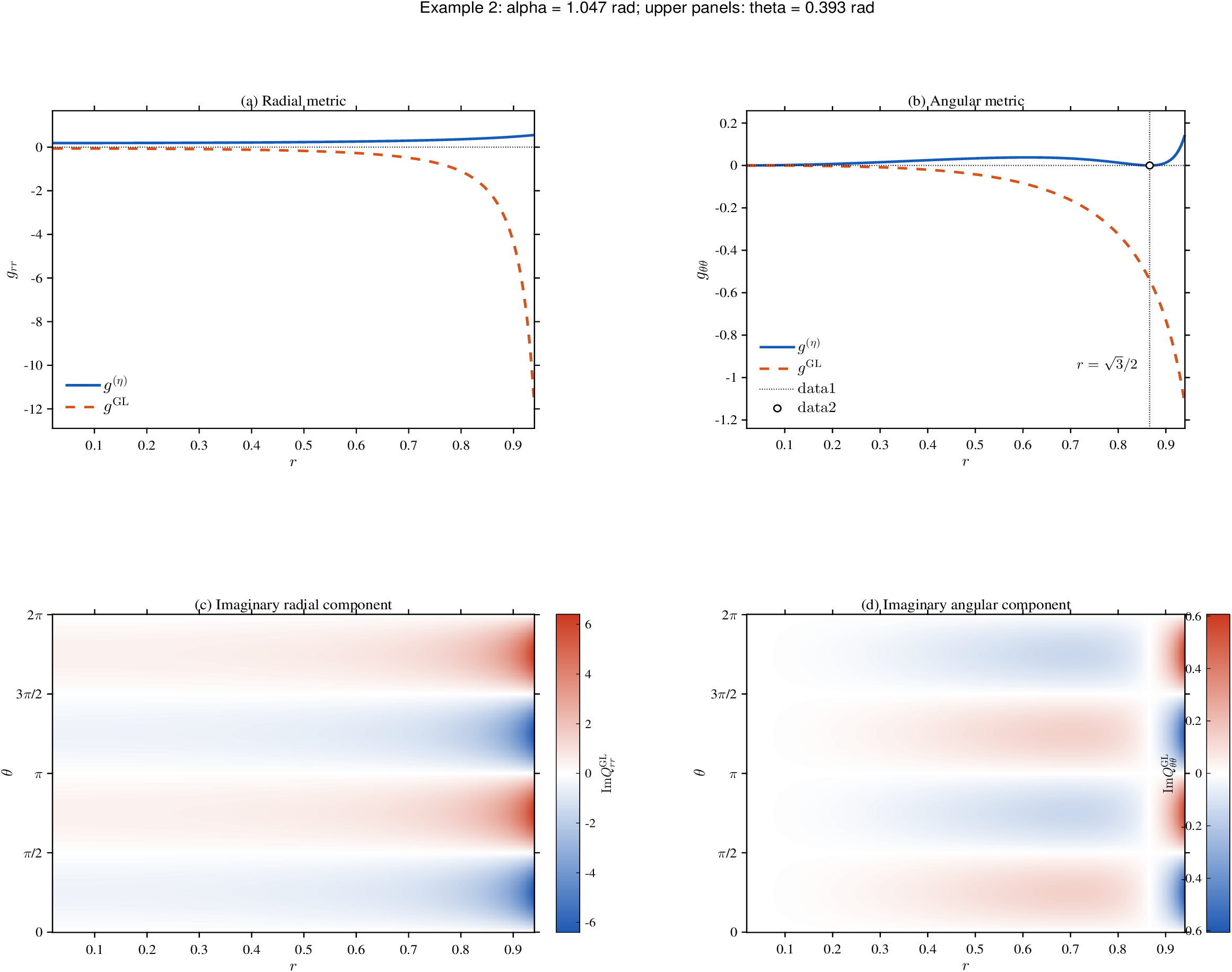}
	\caption{Metrics of the two QGTs and diagonal imaginary parts of Type II for Example 2 with $\alpha=\pi/3$. (a,b) Radial and angular metrics compared at $\theta=\pi/8$. (c,d) $\operatorname{Im}Q_{rr}^{\mathrm{GL}}$ and $\operatorname{Im}Q_{\theta\theta}^{\mathrm{GL}}$, which do not represent the Berry curvature.}
	\label{fig:QGT23}
\end{figure}

\emph{Type II QGT.}
The auxiliary function entering the angular Type-II components is $h(r)=\frac{r}{2\rho}$, with $\rho$ defined above.
A straightforward calculation shows that Eq.~\eqref{QGT_2} can be rewritten as $Q_{\mu\nu}^{\mathrm{GL}}=\mathcal{L}_\mu\mathcal{T}_\nu$  
with
\begin{align}
\mathcal{L}_r
&=\frac{\beta_r'}{2}
-\frac{\cos\beta_r\cos2\theta+\mi\sin2\theta}{2\rho^2},\notag\\
\mathcal{T}_r
&=\frac{\beta_r'}{2}
+\frac{\cos\beta_r\cos2\theta-\mi\sin2\theta}{2\rho^2},\notag\\
\mathcal{L}_\theta
&=h(r)\cos\beta_r\sin2\theta
+\mi\left[\left(g(r)-\frac12\right)\sin\beta_r-h(r)\cos2\theta\right],\notag\\
\mathcal{T}_\theta
&=-h(r)\cos\beta_r\sin2\theta
-\mi\left[\left(g(r)-\frac12\right)\sin\beta_r+h(r)\cos2\theta\right].
\label{ex2-LT}
\end{align}
Specifically, the diagonal components are
\begin{align}
Q_{rr}^{\mathrm{GL}}
=&\frac{(\beta_r')^2}{4}
-\frac{\cos^2\beta_r\cos^22\theta+\sin^22\theta}{4\rho^4}
-\frac{\mi\beta_r'\sin2\theta}{2\rho^2},\notag\\
Q_{\theta\theta}^{\mathrm{GL}}
=&\left(g(r)-\frac12\right)^2\sin^2\beta_r
-h(r)^2\left(\cos^22\theta+\cos^2\beta_r\sin^22\theta\right)
\notag\\
&-2\mi h(r)\left(g(r)-\frac12\right)
\sin\beta_r\cos\beta_r\sin2\theta .
\label{ex2-QGL-diagonal}
\end{align}
Generically $Q_{\mu\nu}^{\mathrm{GL}}\neq(Q_{\nu\mu}^{\mathrm{GL}})^*$, and the diagonal components can be complex.
The induced real symmetric tensor $g^{\mathrm{GL}}$ is obtained from Eq.~\eqref{GL_pseudo_metric_definition}, with components
\begin{align}
g_{rr}^{\mathrm{GL}}
&=
\frac{(\beta_r')^2}{4}
-
\frac{\cos^2\beta_r\cos^22\theta+\sin^22\theta}{4\rho^4},\notag\\
g_{r\theta}^{\mathrm{GL}}
&=
-\frac{h(r)\sin^2\beta_r}{2\rho^2}
\sin2\theta\cos2\theta,\notag\\
g_{\theta\theta}^{\mathrm{GL}}
&=
\left(g(r)-\frac12\right)^2\sin^2\beta_r\notag\\
&\quad-h(r)^2
\left(\cos^22\theta+\cos^2\beta_r\sin^22\theta\right).
\end{align}
These additional real and imaginary parts originate from the full derivative of the physical dual and persist even on a single circle; the full $Q^{\mathrm{GL}}$ is therefore complex and generally non-Hermitian. This example shows that when state and metric variations are intertwined, the diagonal components of $Q^{\mathrm{GL}}$ are generically complex, so the real line element alone is insufficient to capture the full tensor. Figure~\ref{fig:QGT23} shows the radial and angular real metrics together with the two diagonal imaginary parts of the Type II QGT: the upper panels compare the real line elements of the two prescriptions for the same state family, while the lower panels show that the real metric does not capture the full complex tensor.

\emph{Topology.} The parameter space is the disk $0\le r<1$, $0\le\theta<2\pi$, which is contractible. Therefore the pullback of the $U(1)$ principal bundle is trivial, and the first Chern number vanishes,
\begin{equation}
C_2^{(\eta,\Gamma)}=C_2^{\mathrm{GL}}=0.
\end{equation}
The nontrivial \(\theta\)-dependence of the connection does not produce a topological invariant, because the radial coordinate \(r\) is not periodic and the parameter space is not a torus.

\subsection{Example 3: topological model on the sphere}
\label{subex3}

Examples 1 and 2 have contractible parameter spaces, so their bundles are topologically trivial. To exhibit nontrivial topology, we take $\mathcal{M}=S^2$ with coordinates $(\theta,\phi)$ and the Hermitian Hamiltonian
\begin{equation}
H_H=\boldsymbol{n}\cdot\boldsymbol{\sigma},
\quad
|\chi_-(\theta,\phi)\rangle=
\begin{pmatrix}
\sin(\theta/2)\\
-\cos(\theta/2)\me^{\mi\phi}
\end{pmatrix},
\end{equation}
where $\boldsymbol{n}=(\sin\theta\cos\phi,\sin\theta\sin\phi,\cos\theta)$. The state $|\chi_-\rangle$ is the $E=-1$ eigenstate of $H_H$. With the Berry-connection convention $\mathcal{A}_\mu=\mi\langle\chi_-|\partial_\mu\chi_-\rangle$ and the standard orientation $\dif\theta\wedge\dif\phi$, its first Chern number is $C_1^H=+1$. Introducing the metric $\eta(\theta)=\me^{2\alpha\cos\theta\,\sigma_z}$ with $\alpha\in\mathbb{R}$, we have $S(\theta)=\me^{\alpha\cos\theta\,\sigma_z}$, and the associated quasi-Hermitian system is given by $H=S^{-1}H_HS$ and $|\Psi_-\rangle=S^{-1}|\chi_-\rangle$. The state family winds nontrivially, and $H^2(S^2,\mathbb{Z})=\mathbb{Z}$ supports nontrivial topology.

\emph{Type I.} Since $S$ is diagonal, $G_\mu=0$, so $\mathcal{A}^\eta=\mathcal{A}^H$. For the chosen eigenstate, $\mathcal{A}^H_\phi=-\cos^2(\theta/2)$, and hence
\begin{equation}
\mathcal{F}^\eta_{\theta\phi}=\partial_\theta\mathcal{A}^H_\phi
=+\frac12\sin\theta,\qquad
C_1^{(\eta,\Gamma)}=\frac{1}{2\pi}\int_{S^2}\mathcal{F}^\eta=+1.
\end{equation}
The sign follows from the stated Berry-connection convention and the standard orientation of $S^2$. A globally smooth metric does not alter the quantized topology.

\emph{Type II.} With $B_\theta=-\alpha\sin\theta\,\sigma_z$ and $B_\phi=0$,
\begin{equation}
a^{\mathrm{GL}}_\theta
=-\alpha\sin\theta\cos\theta,
\quad
a^{\mathrm{GL}}_\phi
=\mi\cos^2(\theta/2),
\end{equation}
and
\begin{equation}
\mathcal{F}^{\mathrm{GL}}_{\theta\phi}
=+\frac12\sin\theta
=\mathcal{F}^\eta_{\theta\phi},\qquad
C_1^{\mathrm{GL}}=C_1^{(\eta,\Gamma)}=+1.
\end{equation}
The $\alpha$-dependence drops out of the curvature and survives only in the open-path complex phase, whose modulus part is $\operatorname{Im}\gamma^{\mathrm{GL}}=\frac{\alpha}{2}(\cos^2\theta_2-\cos^2\theta_1)$.

The equality of the two Chern numbers is general. By Eq.~(\ref{GL_reduction_connection}),
\begin{equation}
\mathcal{A}^{\mathrm{GL}}_{n,\mu}-\mathcal{A}^\eta_{n,\mu}
=-\frac{\mi}{2}\langle\Psi_n|\partial_\mu\eta|\Psi_n\rangle
\end{equation}
is a globally defined one-form. Consequently, the curvature difference is exact,
\begin{equation}
\mathcal{F}^{\mathrm{GL}}-\mathcal{F}^{\eta}
=\dif(\mathcal{A}^{\mathrm{GL}}-\mathcal{A}^{\eta}),
\end{equation}
and for any closed oriented surface $\Sigma$,
\begin{equation}
C_1^{\mathrm{GL}}-C_1^{(\eta,\Gamma)}
=\frac{1}{2\pi}\int_\Sigma\dif(\mathcal{A}^{\mathrm{GL}}-\mathcal{A}^{\eta})
=\frac{1}{2\pi}\int_{\partial\Sigma}(\mathcal{A}^{\mathrm{GL}}-\mathcal{A}^{\eta})=0.
\end{equation}
Thus the two constructions define the same first Chern class.

\subsection{Comparison of the models}
\label{submodels_comparison}

The three retained figures focus on the QGT comparison: Fig.~\ref{fig:QGT11} illustrates the common proper representation in Example 1, Fig.~\ref{fig:QGT12} compares the two tangent-space metrics in Example 1, and Fig.~\ref{fig:QGT23} compares the radial and angular metrics and the diagonal imaginary components in Example 2.

The models illustrate three points. (i) The distinction between the two QGTs appears at fixed physical metric, state family, and normalized section, so it is not due to normalization. In Example 1 it is isolated in the $u$ direction, where the covariant derivative compensates the state change while the endpoint comparison retains the metric differential, giving $g_{uu}^{\mathrm{GL}}<0$ and signature $(1,1)$; in Example 2 state and metric variations are intertwined, and the diagonal components of $Q^{\mathrm{GL}}$ are generically complex. (ii) In both models $g^{\mathrm{GL}}$ equals $g^{(\eta,\Gamma)}$ minus the horizontal Gram term, Eq.~(\ref{two_metrics_relation}), and $Q^{\mathrm{GL}}$ factorizes as $\mathcal{L}_\mu\mathcal{T}_\nu$ with complex rank at most one, while the real metric can be non-degenerate; complex and real ranks must be judged separately. In Example 1 the symmetric $Q^{\mathrm{GL}}$ has zero curvature, whereas Example 2 has a nonzero antisymmetric part with curvature $\mi(Q^{\mathrm{GL}}_{\mu\nu}-Q^{\mathrm{GL}}_{\nu\mu})$. (iii) Example 3 shows that on $S^2$ both constructions carry the same Chern number, differing only in non-topological open-path complex phases. The two constructions thus organize different local metric and connection information while sharing the same global topology.

\section{Conclusion}\label{Sec6}

We have formulated and compared two quantum geometric tensors for parameter-dependent quasi-Hermitian systems, distinguished by how states at different parameter points are compared. The Type I QGT is constructed from the metric-compatible covariant derivative and the corresponding horizontal projection. It is a Hermitian, positive-semidefinite Gram tensor and provides an intrinsic geometry of state-ray variations relative to the chosen physical connection. For any normalized solution of the inner-product-preserving dynamics, its tangential component is exactly the physical energy variance, establishing a direct connection with the Anandan--Aharonov evolution speed.

The Type II QGT is obtained by differentiating the physical dual in full and expanding a $GL(1,\mathbb{C})$-invariant endpoint overlap. On a normalized section, this tensor is algebraically identical to the LR QGT introduced by Das \emph{et al.}~\cite{Das2026}. Our contribution is to identify its symmetric endpoint-overlap construction and to compare it directly with the independently defined Type I QGT. In the positive Dyson frame, the Type I tensor is the Gram block of the dressed-derivative state, while the full LR/Type II tensor additionally contains the stretching-fluctuation contribution and mixed terms. This relation is expressed by Eq.~\eqref{two_qgt_exact_relation}. In particular, the real symmetric part of Type II is the Type I quantum metric minus a positive-semidefinite term from the metric connection. The resulting indefinite directions therefore arise from the endpoint comparison and metric variation, not from any loss of positivity of the physical Hilbert-space inner product.

We have also clarified the role of proper Dyson mappings in the comparison with the Hermitian $K$-type QGT: a pathwise proper map ensures equality of tangential contractions, whereas equality of all tensor components requires a common proper frame. The three examples illustrate the distinct local geometries of Type I and Type II and the circumstances under which the Dyson-induced and Type I constructions coincide.

The physical interpretations of the two tensors are correspondingly different. Type I controls the energy-uncertainty contribution to the speed of quantum evolution. Type II characterizes the symmetrized endpoint overlap, a complex $GL(1,\mathbb{C})$ holonomy, and a signed quadratic response along the dynamics. These results clarify the distinct geometric information retained by a metric-compatible connection and by a direct endpoint comparison, and provide a unified basis for studying quantum geometry in parameter-dependent quasi-Hermitian systems.
\begin{acknowledgments}
H.G. was supported by the Quantum Science and Technology-National Science and
Technology Major Project (Grant No.~2021ZD0301904) and the National Natural
Science Foundation of China (Grant No.~12447216). X.-Y. H was supported by the National Natural Science Foundation of China (Grant No. 12405008).

\end{acknowledgments}

\appendix
\section{K\"ahler geometry and the QGT--Berry curvature correspondence}
\label{app:kahler}

In Hermitian quantum mechanics, the normalized state space is $S^{2N-1}$, and the physical state space is the projective space $\mathbb{CP}^{N-1}=S^{2N-1}/U(1)$, which is the base of the Hopf fibration. The Berry connection and curvature on this $U(1)$ principal bundle are naturally tied to the K\"ahler geometry of $\mathbb{CP}^{N-1}$. In this appendix we explain why this identification holds in the Hermitian case, and why the imaginary part of the Type I QGT need not coincide with the Berry curvature when the physical metric depends on parameters.

\subsection{Why the complex structure ties the real and imaginary parts together}

The QGT in Eq.~\eqref{Qmn} is a complex inner product of horizontal vectors, $Q_{\mu\nu}
=
\langle h_\mu|h_\nu\rangle$ with $|h_\mu\rangle
=
P^\perp|\partial_\mu\psi\rangle$.
Thus, its real and imaginary parts are given by $g_{\mu\nu}
=
\operatorname{Re}\langle h_\mu|h_\nu\rangle$ and $\Omega_{\mu\nu}
=
-\operatorname{Im}\langle h_\mu|h_\nu\rangle$ respectively.
On $\mathbb{CP}^{N-1}$, there is a natural complex structure $J$ acting on horizontal vectors by multiplication by $\mi$,
\begin{equation}
J|h_\mu\rangle=\mi|h_\mu\rangle.
\end{equation}
This operator satisfies $J^2=-1$ and is compatible with the Hermitian inner product. Its presence is what ties $g_{\mu\nu}$ and $\Omega_{\mu\nu}$ together. Indeed, consider
\begin{equation}
\langle Jh_\mu|h_\nu\rangle
=
\langle \mi h_\mu|h_\nu\rangle
=
-\mi\langle h_\mu|h_\nu\rangle.
\end{equation}
Taking the real part,
\begin{equation}
\operatorname{Re}\langle Jh_\mu|h_\nu\rangle
=
\operatorname{Re}\bigl(-\mi\langle h_\mu|h_\nu\rangle\bigr)
=
\operatorname{Im}\langle h_\mu|h_\nu\rangle
=
-\Omega_{\mu\nu}.
\end{equation}
Therefore,
\begin{equation}
\Omega_{\mu\nu}
=
-\operatorname{Re}\langle Jh_\mu|h_\nu\rangle
=
-g(Jh_\mu,h_\nu),
\end{equation}
where $g(\cdot,\cdot)$ denotes the real inner product. Equivalently,
\begin{equation}
\Omega(X,Y)=g(X,JY)=-g(JX,Y).
\end{equation}
This is the compatibility relation among the Fubini--Study metric, the complex structure, and the associated two-form. With the convention used here, the two-form $\Omega$ is the K\"ahler form $\omega$. On $\mathbb{CP}^{N-1}$, the standard complex structure is integrable and the corresponding two-form is closed, so these structures define its standard K\"ahler geometry.

The geometric reason is that on a K\"ahler manifold the Hermitian metric $h$, the Riemannian metric $g=\operatorname{Re}h$, and the K\"ahler form $\omega=-\operatorname{Im}h$ are three aspects of the same geometric structure. The complex structure $J$ relates them by
\begin{equation}
\begin{aligned}
\omega(X,Y)&=g(X,JY)=-g(JX,Y),\\
g(X,Y)&=\omega(JX,Y).
\end{aligned}
\end{equation}
Thus, once $g$ is known, $\omega$ is determined by $J$, and vice versa. This compatibility concerns the tangent geometry of the projective Hilbert space; it does not by itself equip an arbitrary external parameter manifold with a compatible complex structure.

\subsection{Why the K\"ahler form is the $U(1)$ gauge curvature}

The Berry connection on the Hopf bundle is $\mathcal{A}_\mu
=
\mi\langle\psi|\partial_\mu\psi\rangle$,
and its curvature is
\begin{equation}
\mathcal{F}_{\mu\nu}
=
\partial_\mu\mathcal{A}_\nu-\partial_\nu\mathcal{A}_\mu
=
-2\operatorname{Im}Q_{\mu\nu}
=
2\Omega_{\mu\nu}.
\end{equation}
The relation follows from the K\"ahler reduction that constructs $\mathbb{CP}^{N-1}$ from $\mathbb{C}^N$. Let $\pi:S^{2N-1}\to\mathbb{CP}^{N-1}$ be the Hopf projection, and let $\mathcal{A}_{\mathrm{Hopf}}=\mi\langle\psi|\dif\psi\rangle$ be the real-valued Hopf connection. Its curvature $\mathcal{F}_{\mathrm{Hopf}}=\dif\mathcal{A}_{\mathrm{Hopf}}$ is horizontal and $U(1)$-invariant, and hence is a basic two-form on $S^{2N-1}$. It therefore descends to the quotient, with the normalization fixed by
\begin{equation}
\pi^*\omega_{\mathrm{FS}}
=
\frac{1}{2}\mathcal{F}_{\mathrm{Hopf}}.
\end{equation}
Equivalently, the pullback of the Fubini--Study K\"ahler form is one half of the curvature of the Hopf connection. Along a state map, this gives $\mathcal{F}=2\omega_{\mathrm{FS}}$ in the convention used above.

In this sense, the K\"ahler form and the $U(1)$ gauge curvature are the same object viewed from two sides.

The physical content of the K\"ahler condition in the conventional Hermitian construction is that the quantum metric $g$ and the Berry curvature $\mathcal{F}$ are not independent: the former is the real part of the QGT, while the latter is twice the associated K\"ahler form, $\mathcal{F}=2\omega=-2\operatorname{Im}Q$. Both are tied together by the complex structure of $\mathbb{CP}^{N-1}$. This identification uses the ordinary differential of the state map and need not extend unchanged to the covariant derivatives used in the Type I construction.

\subsection{Why pointwise K\"ahler geometry need not extend to the Type I parameter-space geometry}

In the quasi-Hermitian setting, the normalized states at each parameter value form $\mathcal{S}_{\eta(\boldsymbol{\lambda})}$, and the corresponding physical rays form $\mathcal{R}_{\eta(\boldsymbol{\lambda})}=\mathcal{S}_{\eta(\boldsymbol{\lambda})}/U(1)$. For every fixed $\boldsymbol{\lambda}$, this ray space is biholomorphic to $\mathbb{CP}^{N-1}$ and carries its natural K\"ahler structure. The biholomorphism is induced by the invertible complex-linear map $|\Psi\rangle\mapsto\eta^{1/2}(\boldsymbol{\lambda})|\Psi\rangle$. Hence, the pointwise complex structure is well defined at every parameter value and is compatible with the corresponding positive-definite metric.

The Type I QGT defines a Hermitian tensor on the external parameter manifold $\mathcal{M}$ through the covariant horizontal map $X\mapsto |h_X\rangle=P_\eta^\perp D_X^\eta|\Psi\rangle$. A K\"ahler structure induced in this way would require the real part of the Type I QGT to be nondegenerate and positive definite, invariance of the image of the horizontal map under multiplication by $\mi$, and an integrable complex structure with a closed associated two-form. In general, the image of the horizontal map need not be invariant under multiplication by $\mi$, so the pointwise complex structure need not descend to $T\mathcal{M}$. Even when an induced almost-complex structure exists, its integrability and the closure of the corresponding two-form require additional conditions. Thus, the pointwise K\"ahler geometries do not by themselves ensure a K\"ahler structure for the parameter-space geometry defined by the Type I QGT.

The Berry curvature associated with the metric-compatible connection is related to the Type I QGT by Eq.~(\ref{berry_ambient_relation}):
\[
\mathcal{F}_{n,\mu\nu}^\eta
=-2\operatorname{Im}Q_{n,\mu\nu}^{(\eta,\Gamma)}
+\mi\langle\Psi_n|\eta\mathcal{R}_{\mu\nu}^\eta|\Psi_n\rangle.
\]
The additional contribution is determined by the curvature $\mathcal{R}_{\mu\nu}^\eta$ of the metric-compatible comparison connection and encodes the effect of parallel transport between parameter-dependent inner-product spaces.

\subsection{Bundle structure, distance decomposition, and parallel transport}

The principal-bundle structure underlying the Type I construction is preserved. For a chosen smooth ray family over $\mathcal{M}$ (an eigen-ray family in the spectral setting), its normalized representatives form a $U(1)$ principal bundle, and the distance decomposition of Sec.~\ref{Sec3b} remains valid. Parallel transport is characterized by the vanishing of the fiber component in this decomposition, with Berry phase $\gamma_n^\eta(C)=\mathlarger{\oint}_C\mathcal{A}_n^\eta$. The associated line bundle has a first Chern class in $H^2(\mathcal{M},\mathbb{Z})$, represented in de Rham cohomology by $\mathcal{F}_n^\eta/(2\pi)$; its integral characterization is given in Sec.~\ref{Sec3b}.

\bibliography{Review4,Review,Review1,Review2,Review3,ProperHermitianization2026}
\end{document}